\documentclass[10pt,journal,final]{IEEEtran}

\def\BibTeX{{\rm B\kern-.05em{\sc i\kern-.025em b}\kern-.08em
    T\kern-.1667em\lower.7ex\hbox{E}\kern-.125emX}}
    
\usepackage[english]{babel}

\usepackage[letterpaper,top=2cm,bottom=2cm,left=3cm,right=3cm,marginparwidth=1.75cm]{geometry}

\usepackage{cite}

\usepackage{booktabs}
\usepackage[table,xcdraw]{xcolor}
\usepackage{newtxmath}
\usepackage{dsfont}
\usepackage{amsmath}
\usepackage{graphicx}
\usepackage[colorlinks=true, allcolors=blue]{hyperref}
\usepackage[figuresright]{rotating}

\usepackage{mdwmath}
\usepackage[switch]{lineno}
\usepackage{adjustbox}
\usepackage{xcolor}
\usepackage[normalem]{ulem}
\usepackage[T1]{fontenc}

\usepackage{adjustbox}
\usepackage{multirow} %multirow for format of table 

\newcommand{\ie}[1]{\textit{i.e.} #1}
\newcommand{\eg}[1]{\textit{e.g.} #1}

\newcommand{\ql}[1]{\textcolor{black}{#1}}
\newcommand{\wk}[1]{\textcolor{black}{#1}}
\newcommand{\zc}[1]{\textcolor{black}{#1}}

\newcommand{\re}[1]{\textcolor{black}{#1}}
\newcommand{\ree}[1]{\textcolor{black}{#1}}

\title{%Opt1: Cortical thickness network analysis in first episode schizophrenia \\
Network analysis on cortical morphometry in first-episode schizophrenia\\
}

\author{Mowen Yin, Weikai Huang, Zhichao Liang,
Quanying Liu$^{\ast}$, Xiaoying Tang$^{\ast}$
\thanks{This study is supported by the National Natural Science Foundation of China (62071210, 62001205), the Shenzhen Science and Technology Program (RCYX20210609103056042, 20200925155957004, KCXFZ2020122117340001), the Shenzhen Basic Research Program (JCYJ20200925153847004, JCYJ20190809120205578), the Guangdong Natural Science Foundation Joint Fund (2019A1515111038), the Shenzhen Key Laboratory of Smart Healthcare Engineering (ZDSYS20200811144003009).}% <-this % stops a space
\thanks{M. Yin, Z. Liang and
Q. Liu are with Shenzhen Key Laboratory of Smart Healthcare Engineering, Department of Biomedical Engineering, Southern University of Science and Technology. W. Huang and X. Tang are with Department of Electronic and Electrical Engineering, Southern University of Science and Technology, Shenzhen 518055, P.R. China}
\thanks{$^{\ast}$ Corresponding to {\tt\small tangxy@sustech.edu.cn} and {\tt\small liuqy@sustech.edu.cn}}
}

\begin{document}

\maketitle

\begin{abstract}
%The neurological decline of first-episode schizophrenia (FES) is closely related to changes in the network topology of the cerebral cortex. 
First-episode schizophrenia (FES) results in abnormality of brain connectivity at different levels.
\re{Despite some successful findings on functional and structural connectivity of FES, relatively few studies have been focused on morphological connectivity, which may provide a potential biomarker for FES.}
In this study, we \ql{aim to} investigate cortical morphological connectivity in FES. T1-weighted magnetic resonance image data from 92 FES patients and 106 healthy controls (HCs) are analyzed.
\re{We parcellate brain into 68 cortical regions, calculate the averaged thickness and surface area of each region, construct undirected networks by correlating cortical thickness or surface area measures across 68 regions for each group, and finally compute a variety of network-related topology characteristics.}
Our experimental results show that both the cortical thickness network and the surface area network in two groups are small-world networks; that is, those networks have high clustering coefficients and low characteristic path lengths. 
At certain network sparsity levels, both the cortical thickness network and the surface area network of FES have significantly lower clustering coefficients and local efficiencies than those of HC, indicating FES-related abnormalities in local connectivity and small-worldness. These abnormalities mainly involve the frontal, parietal, and temporal lobes. Further regional analyses confirm significant group differences in the node betweenness of the posterior cingulate gyrus for both the cortical thickness network and the surface area network. %done\todo{I did not get it}. %For surface area network, the entorhinal cortex also reflected significant differences in the distance between nodes.Our results provide new evidence for FES-associated brain network abnormalities.
%\ql{We provide evidence of cortical morphological network abnormalities in FES, and the network topology characteristics can be effective biomarkers of FES}.
%Our work support that constructing cortical morphological connectivity using correlation of cortical thickness across subjects can reflect topological abnormality of the neurology disorders. 
\re{Our work supports that cortical morphological connectivity\wk{, which is constructed based on correlations across subjects' cortical thickness,} may serve as a tool to study topological abnormalities in neurological disorders.}
\end{abstract}


\section{Introduction}

%Schizophrenia is a kind of chronic mental condition with clinical manifestations of hallucinations, negative and cognitive deficit symptoms for which the etiology is not yet clear. 
Schizophrenia is a chronic psychiatric disorder characterized by hallucinations, negative symptoms, and cognitive deficits. First-episode schizophrenia (FES) is an early phase of schizophrenia when an individual first presents with schizophrenia-consistent symptoms and formally receives a definitive diagnosis of schizophrenia after professional evaluations\cite{wiersma1998natural}. This phase usually happens in teenagers or early twenties. Studies on FES can help identify different subtypes of schizophrenia from disease evolution perspective and help suggest effective treatment plans ~\cite{larsen199601,weiden2007understanding,murray1997global}.

Advanced neuroimaging techniques (\eg magnetic resonance imaging (MRI)~\cite{MRI,Neuroimaging}, computer tomography (CT)~\cite{CT,Neuroimaging}, and positron emission tomography (PET)~\cite{PET,Neuroimaging}) have been applied to diagnose and analyze schizophrenia.
\re{Among them, MRI is widely used due to} its multiple imaging directions, high spatial resolution and accurate positioning. \re{MRI has the potential to be beneficial for studying the FES-related morphological abnormality patterns of the human brain}.
Previously reported FES-related brain abnormalities, as revealed by MRI, mainly focus on specific brain regions such as the amygdala, hippocampus and corpus callosum\cite{Tang2020,1Tang2021}, cognitive deterioration\cite{ADDINGTON199850} and so on. However, \re{it might be wrong to treat each brain region as a separate object, as} the human brain is a complex network in which multiple regions work in coordination~\cite{Sporns2004015}. 
Different brain regions are interactive and coordinated in terms of structural, functional and morphological connectivity\cite{sporns2016networks}. Such a networked organization facilitates the specialization and integration of various brain functions, enabling efficient and complex functional activities of the brain~\cite{Sophie2007013,Bullmore2009012,liu2017detecting}. The occurrence of a specific mental disease is often accompanied by abnormal brain network topology. In recent years, studies have revealed FES-related abnormalities in terms of complex brain networks that involve abnormal interactions among various brain regions~\cite{Bassett200907,ZHU2012611}. In brain network analysis, there are typically three types of networks: (i) functional connectivity network (FCN)\cite{LynnFCNCMN,VANDENHEUVEL2010519}, (ii) structural connectivity network (SCN)\cite{ZHANG2012109,LifeiDTI}, and (iii) morphological connectivity network (MCN)\cite{LynnFCNCMN,ZUGMAN201589}. Previous network analysis studies on FES are mainly focused on FCN and SCN, but MCN is relatively less investigated. 

\re{Cortical morphological network (CMN), as a specific type of MCN, \wk{focuses} on morphological characteristics of cortical regions\cite{LiCMN2020} \wk{and} has received increasing attention in FES studies\cite{WANG2021179,JiangFES2021}.}
% 分
Previous CMN-based FES studies have reported preliminary evidence of topological abnormalities in the cortical regions. For example, patients with FES have high CMN modularity, degree, and betweenness centrality in the medial orbitofrontal, fusiform, and superior frontal gyri~\cite{WANG2021179}. Jiang et al. suggests that FES patients have increased cortical covariance between regions with thinner cortical thickness such as the prefrontal and temporal lobe regions, compared with a healthy control (HC) group\cite{JiangFES2021}. Although there is sporadic evidence of changes in the network properties of CMNs in FES, more studies are needed to provide convergent and comprehensive evidence.
%These findings suggest that the morphometry network of schizophrenic patients has less ideal topological organization, especially degradation of small-world properties. These topological properties not only provide a new perspective for understanding the pathological mechanism of schizophrenia by studying the differences in the topological properties of the brain morphometry network, but also can be used as biological markers and therapeutic targets of the disease.

%This degradation of small-world attributes means that the global and local efficiency of the brain network in processing information is reduced. By analyzing the network topology characteristics of the brains of patients with schizophrenia, and judging whether there are similar abnormalities, the universality of these abnormal characteristics is confirmed.

\re{In this study, we aim to investigate the network properties of two cortical morphological networks (\ie a cortical thickness network and a surface area network) in patients with FES, compared with healthy controls.} Specifically, we collect 3D T1-weighted MRI data from 92 FES patients and 106 HCs. We first obtain 68 cortical regions, calculate the average thickness and surface area of each region, and then construct 68-node undirected networks by correlating cortical thickness or surface area measures across 68 cortical regions for each group. In order to provide a comprehensive understanding of the FES-related abnormality in CMN organizations, a variety of network-related topology characteristics are computed and compared between FES group and HC group, including the global and local efficiency, small-worldness, characteristic path length, and betweenness centrality.
Our experimental results demonstrate small-worldness in both cortical thickness network and surface area network of the two groups; that is, those networks have high clustering coefficients and low characteristic path lengths. 
Both the thickness network and the surface area network of FES have significantly lower aggregation coefficients and local efficiencies than those of HC at certain network sparsity levels (\re{$p<0.05$}), indicating FES-related abnormalities in local connectivity and small-worldness. These abnormalities of FES group mainly involve the frontal, parietal, and temporal lobes. Further regional analyses confirm significant group differences in the node betweenness of the posterior cingulate gyrus (PCC) for both the thickness network and the surface area network.

\section{Materials and Methods}

\subsection{Subjects}
% revised by Quanying
In this study, we recruit 198 subjects in total, including 92 FES patients (age: 22.40 ± 5.59) and 106 healthy subjects (age: 23.68 ± 4.04). The clinical and demographic characteristics of the enrolled subjects are shown in \textbf{Table}~\ref{tab:subjects}.
There is no significant group difference in age ($p = 0.071$), nor gender distribution ($p = 0.16$) between the two groups \re{by permutation test}. 
The research protocol has been approved by the First Affiliated Hospital of Shenzhen University. The FES patients are examined with the DSM-IV\cite{DSM-IV} criteria to confirm a consensus diagnosis of schizophrenia before MRI examination. The \re{exclusion} criteria include history of drug dependence, pregnancy, other diseases of the central nervous system, as well as unstable medical conditions. In order to eliminate the confounding effect of neuroleptic medication, all FES patients are antipsychoticnaïve. All subjects receive MRI scanning in the First Affiliated Hospital of Shenzhen University, after signing the informed consent.

\begin{table}[!t]
\centering
\caption{Clinical and demographic characteristics of subjects.}
\scalebox{0.75}{
\begin{tabular}{lcc}
\hline
\textbf{Characteristic} & \textbf{FES group}  & \textbf{HC group} \\ 
\hline
Number of subjects & 92 & 106 \\
Gender(male/female) & 42/50 & 59/47 \\
Age(years) & 22.40 ± 5.59 & 23.68 ± 4.04 \\
Duration of education (years) &12.32 ± 3.18 &	12.76 ± 3.29 \\
\hline
Age of onset (years)&	21.26 ± 5.29&	$\backslash$ \\
Duration of untreated psychosis (months)&	14.66 ± 23.43&	$\backslash$ \\
GAF scores&	29.16 ± 10.18&	$\backslash$ \\
\hline
\textbf{PANSS scores}:\\		
$\hspace{4mm}$Total&	93.90 ± 16.24&	$\backslash$ \\
$\hspace{4mm}$Positive&	24.55 ± 6.31&	$\backslash$ \\
$\hspace{4mm}$Negative&	20.89 ± 8.46&	$\backslash$ \\
$\hspace{4mm}$General psychopathology&	48.46 ± 8.39&	$\backslash$ \\
\hline
\end{tabular}
}
\label{tab:subjects}
\end{table}

\subsection{MRI Data Acquisition and Preprocessing}
All subjects keep their eyes open during the scan and passively stared at the \re{central cross to maintain immobility}. A 3-Tesla Siemens triple scanner collects all morphometry MRI data, and each subject collects T1-weighted 3D volume images of the entire brain. The image is gathered using a magnetization prepared-rapid acquisition gradient echo (MPRAGE) sequence, \zc{with scanning parameters} repetition time = $13.40$ ms, echo time $= 4.6$ ms, flip angle $= 20$, the field of view (FOV) $= 256 \times 256$, and the isotropic resolution of the entire skull is 1 mm$^3$. A neuroradiologist performs a visual inspection of all MR images to control data quality.

The workflow of our study is in \textbf{Figure} \ref{fig:pipeline}. The MRI data is pre-processed using Freesurfer software package (\url{http://surfer.nmr.mgh.harvard.edu/}). After co-registration, all data is then imported into Freesurfer to extract the boundary between the gray matter and white matter of each cerebral hemisphere, as well as the gray matter endothelial layer. To complete the three-dimensional reconstruction of the brain area, we process the data with the following steps, including mgz data format conversion, non-brain structure removal, volume data registration, white matter separation, local area correction, smoothing, cortical reconstruction, and segmenting regions of interests (ROIs) using Desikan-Killiany atlas. Meanwhile, we export the cortical thickness and surface area parameters of each region.

\ree{To characterize more detailed abnormality patterns of the human brain's network topology, we conduct experiments with a more refined atlas, the Brainnetome atlas~\cite{JIANG2013263}. The Brainnetome atlas contains 210 brain regions, which is much more refined than the Desikan-Killiany atlas. All other analyses remain the same.}
%We first map the Brainnetome atlas to the preprocessed MRI, and then parcellate the individual MRI into 210 brain regions, and build the cortical thickness network using Pearson correlation across subjects. We replicate the graph theoretical network analysis on the cortical thickness network.}
% Then transforms the data into mgz format, removes non-brain structure, and divides the white matter , Local area correction, smoothing, cortical reconstruction, Desikan-Killiany map segmentation and other methods to finally complete the three-dimensional reconstruction of the brain area. At the same time, a text file of cortical thickness parameters of each region is generated under the relevant files.

\subsection{Graph Construction using Cortical Thickness and Surface area}

\begin{figure*}[!htpb]
    \centering
    \includegraphics[scale=0.5]{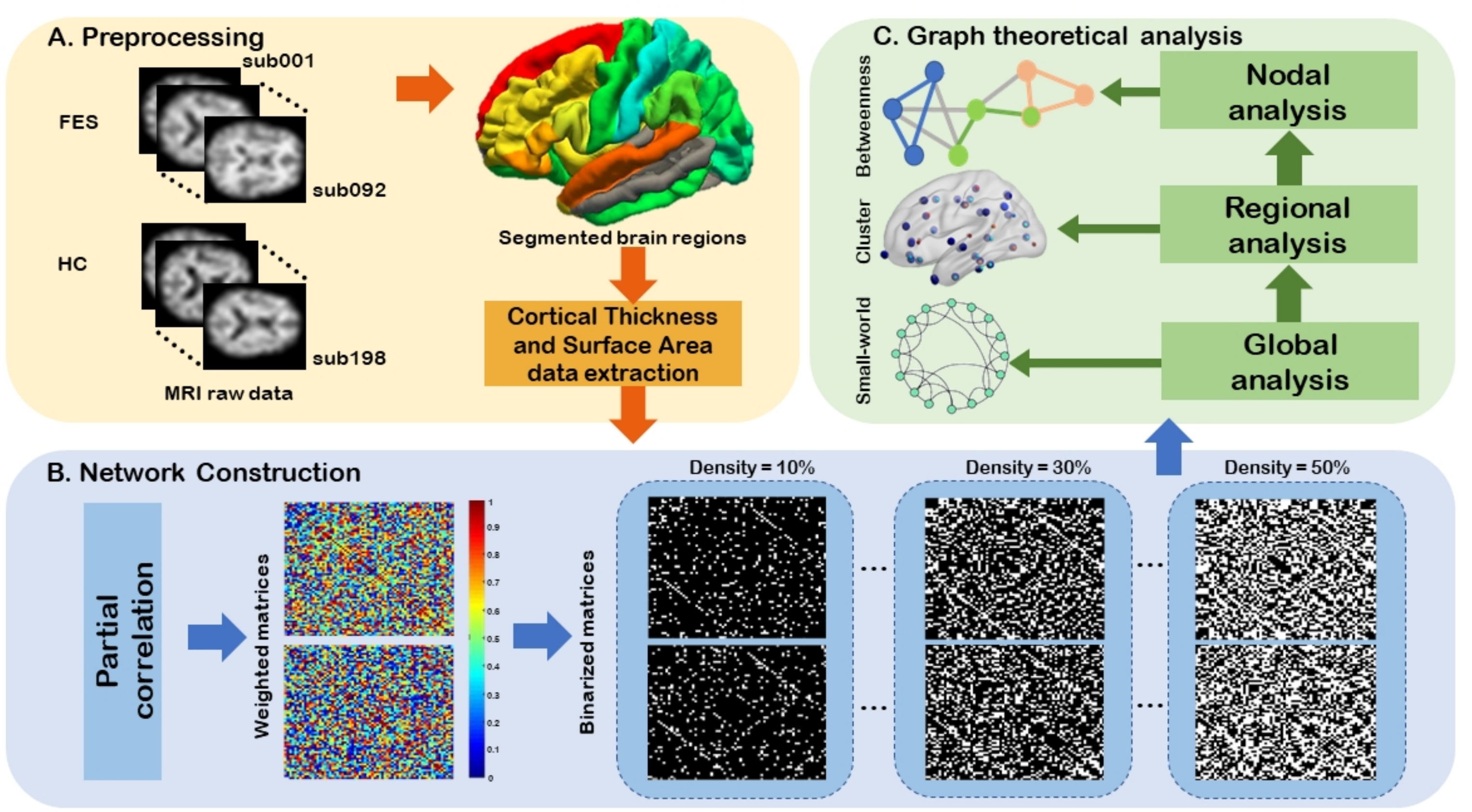}
    \caption{The pipeline for graph theoretical analysis in FES cortical morphological networks. (a) loading MRI raw data of the FES group and HC group. Pre-procession of MRI data and region segmentation into DK68. (b) Construct weighted correlation matrices of cortical thickness after elimination of covariates. Binary correlation matrix according to different density thresholds. (c) Graph theory analysis at different scales.}
    \label{fig:pipeline}
\end{figure*}

As \textbf{Figure} \ref{fig:pipeline} shows, we first examine the differences in cortical thickness between FES and HC groups, and then we represent anatomical connections in terms of the statistical correlation of cortical thickness between brain regions. We divide the brain into 68 regions (the generated morphological matrix is $68 \times 68$) by using the Desikan-Killiany atlas ~\cite{DESIKAN2006968}, \re{which is a gyral-based atlas widely used in brain morphological studies~\cite{POTVIN201743,DK2JAO}}. % By using the Desikan-Killiany atlas template the brain is divided into 68 regions, so the size of the generation matrix is $N \times N$, where $N = 68$). 
Linear regression is performed on the cerebral cortex thickness data of each subject to eliminate the influence of covariates (age, gender), and replace the original residual morphological values for the next application. The partial correlation coefficient of 198 subjects is calculated to generate an inter-region correlation matrix. \re{We apply partial correlation analysis instead of Pearson analysis to remove the effect of other variables since the sample size is relatively large ($>$50).} To test the statistical significance ($p$-value) of a large number of correlation analyses, $p$-values are corrected for multiple comparisons by controlling the family-wise error rate (FWER) at a level of $p \leq 0.05$.
% Because of the large number of correlation analyses, in order to test the significance of these correlations, it is necessary to make multiple corrections. We applied a family-wise error rate (FWER) to correct multiple comparisons with a q value of 0.05. 
The \zc{statistical significant} brain regions are listed in \wk{\textbf{Table}~\ref{tab:ROIs}}, and their three-dimensional performance is shown in \wk{\textbf{Figure}~\ref{fig:thickness}. We perform \re{Fisher's} r to z transformation to transform the correlation matrix into a binary connection matrix and take the sparsity of the connection matrix as the threshold.} 
Using this threshold, if the cortical thickness between the two areas \zc{is statistically correlated}, its element is 1; otherwise, it is 0. Thus, a morphological magnetic resonance connection network of the brain is constructed.

\subsection{Graph theoretical analysis}

The morphological connectivity matrix is denoted as $G \in \mathds{R}^{N \times N}$,

where $N$ is the number of nodes in the network. The subgraph $G_i$ is described as the directly adjacent node set of the $i$-th node. The degree of each node $K_i$ ($i = 1, 2, \dots, 68$) represents the number of nodes in the corresponding subgraph $G_i$, \re{and also present to be the number of edges incident upon it.} The connectivity $K_{net}$ reflects the sparseness of the network:

\begin{equation}\label{eq:K-net}
K_{net} = \frac{1}{N} \sum_{i \in G} K_i .
\end{equation}

The network connectivity density ($K_{cost}$) of a network with $N$ nodes, is defined as the ratio of the actual number of edges in the network to the maximum possible number of edges:

\begin{equation}\label{eq:K-cost}
K_{cost} = \frac{1}{N(N-1)} \sum_{i \in G} K_i .
\end{equation}

\subsubsection{Global network properties}

The \textbf{clustering coefficient} of a node ($C_i$) refers to the proportion of the number of edges actually existing between the nodes directly connected to the node $i$ in the network $E_i$ to the maximum number of edges.
\begin{equation}\label{eq:C-i}
C_{i} = \frac{E_i}{K_i (K_i-1)/2}.
\end{equation}

The \textbf{network average clustering coefficient} ($C_{net}$) is the mean of clustering coefficients across all nodes in the whole network\re{, representing} the degree to which nodes in a graph tend to cluster together. 
\begin{equation}\label{eq:C-net}
C_{net} = \frac{1}{N} \sum_{i \in G} C_i .
\end{equation}

\re{Similar to the clustering coefficient, the \textbf{transitivity} ($T$) of a network also measures the tendency of nodes to cluster together. High transitivity means that the network contains internally densely connected communities or groups of nodes.}
\re{
\begin{equation}\label{eq:T}
T = \frac{3 \times n_{triangles}}{n_{triples}} .
\end{equation}
}

\re{Where $n_{triangles}$ is defined as the number of triangles in the network, $n_{triples}$ \wk{is the} number of connected triples of vertices.}

The \textbf{shortest distance} between node pairs is defined as the minimum number of edges from node $i$ to node $j$. $L_{ij}$ is the distance from node $i$ to node $j$. When the two vertices are not connected, the distance is infinity. The \textbf{characteristic path length} of the network $L_i$ is defined as the average of the shortest distance between all pairs of nodes. 
\begin{equation}\label{eq:L-i}
L_{i} = \frac{1}{N-1} \sum_{i \neq j \in G} \text{min}L_{ij} .
\end{equation}

The \textbf{average path length} of the whole network ($L_{net}$) refers to a quantitative indicator of the tightness of the entire network.
\begin{equation}\label{eq:L-net}
L_{net} = \frac{1}{N} \sum_{i \in G} L_{i} .
\end{equation}

\re{To examine the small-world property of networks, we compare the cortical morphological networks with the constructed regular network and random network. A regular network has only short connected edges, so it consumes fewer resources, while a random network has many long connected edges, which consume more resources.} Here, $L_{net}^{real}$ and $C_{net}^{real}$ denote the characteristic path length and the clustering coefficient of the regular network, respectively. $L_{net}^{random}$ and $C_{net}^{random}$ denote the characteristic path length and the clustering coefficient of the random network, respectively. A network that has the following two properties is considered a small-world network:

\begin{equation}  
\left\{  
    \begin{array}{lcc}  
       \gamma > 1,  &  \text{where} & \gamma =  C_{net}^{real} / C_{net}^{random} ; \\  
       \lambda \sim 1,  & \text{where} & \lambda = L_{net}^{real} / L_{net}^{random} .
    \end{array}  
\right.
\end{equation} 

The parameter $\sigma = \gamma / \lambda$ is defined to quantify the small-world attributes of the network. $\gamma$ \re{is defined as normalized clustering coefficient,} reflecting the local information processing efficiency of the network. The higher the $\gamma$, the stronger the network's ability to process information locally. $\lambda$ \re{is normalized path length}, which mainly reflects the information transmission and the integration efficiency of the network. The shorter the $\lambda$, the stronger the network's overall ability to process information\cite{Bullmore2009012}.

The \textbf{global efficiency} $E_{global}$ quantifies the exchange of information across the whole network where information is concurrently exchanged. For a given graph $G$ with $N$ nodes, $E_{global}$ is described as the average of the reciprocal of the shortest path between each two nodes:
\begin{equation}\label{eq:E-global}
E_{global} = \dfrac{1}{N(N-1)} \sum_{i \neq j \in G} \dfrac{1}{L_{ij}}.
\end{equation}

The \textbf{local efficiency} $E_{local}$ characterizes how well information is exchanged by its neighbors when it is removed. \re{The local efficiency of a node $i$ is:
\begin{equation}\label{eq:E-local-i}
E_{local}(i) = \dfrac{1}{N(N-1)} \sum_{i \neq j \in G} \dfrac{1}{L_{ij}}.
\end{equation}}

\re{The local efficiency of the network is the averaged local efficiency across all nodes. It is defined as:
\begin{equation}\label{eq:E-local}
E_{local} = \dfrac{1}{N} \sum_{i \in G} E_{local}(i).
\end{equation}
}

The brain network can be considered as a small-world network as it meets two criteria as follows,
\begin{itemize}
    \item $E_{global}(G_{regular}) < E_{global}(G_{real}) < E_{global}(G_{random})$,
    \item $E_{local}(G_{regular}) < E_{local}(G_{real}) < E_{local}(G_{random})$,
\end{itemize}
where $E_{global}(G_{regular})$, $E_{global}(G_{real})$ and $E_{global}(G_{random})$ represent the global efficiency of the regular network, the real network and the random network, respectively. $E_{local}(G_{regular})$, $E_{local}(G_{real})$ and $E_{local}(G_{random})$ represent the local efficiency of the regular network, real network and random network, respectively.

\textbf{Betweenness centrality} characterizes how often a node or edge lies on the shortest paths between all pairs of nodes. \textbf{Node betweenness} $B_{node}$ is defined as the ratio of the number of all shortest paths through bthe node to the total number of shortest paths in the network.
\begin{equation}\label{eq:B-node}
B_{node}(i) = \sum_{j,k \in G} \dfrac{\mu_{jk}(i)}{\mu_{jk}},
\end{equation}
where $\mu_{jk}$ is the total number of \re{the} shortest path between node $j$ and node $k$. $\mu_{jk}(i)$ is number of those paths that pass through $i$.

\textbf{Edge betweenness} is described as the ratio of the number of all shortest paths through the edge to the total number of shortest paths in the network.
\re{
\begin{equation}\label{eq:B-edge}
B_{edge}(m) = \sum_{j,k \in G} \dfrac{\mu_{jk}(m \in l)}{\mu_{jk}}.
\end{equation}
}
    \re{where $l$ represents the shortest paths between node $j$ and $k$ and $\mu_{jk}(m \in l)$ denotes} the number of shortest paths between node $j$ and node $k$ passing through \re{edge $m$.}

\subsubsection{Modularity}

\textbf{Modularity} is a measure of network separation. \re{High modularity means that the internal link density of the network is high, and low modularity means sparse.} The advantage of a modular system is that it can evolve by \re{changing} one module at a time without the risk of losing \re{the functionality} of other modules that are well adapted\cite{MeunierModular}. Modularity is defined as:
\re{
\begin{equation}\label{eq:Modul}
Q = \dfrac{1}{2M}\sum_{ij} (A_{ij}-\dfrac{K_iK_j}{2M})\delta(c_i,c_j).
\end{equation}  }
\re{where $M$ is the total number of edges in network and node $i$ belongs to community $c_i$. $A_{ij}$ is an element of the Adjacent matrix of the network thus $A_{ij} =1$ if node $i$ and $j$ are connected, and
$A_{ij} =0$ otherwise.} \re{Besides, the $\delta$-function $\delta(c_i,c_j)$ is 1 if $c_i = c_j$ and 0 otherwise.} The larger the $Q$ value is, the more obvious the community structure in the network is.

\subsubsection{Assortativity}
\textbf{Assortativity} is a measure of the relationship between pairs of connected nodes.% The Assortativity coefficient resulting $r$ represents a positive value means a certain degree of synergy between the same points, while a negative value indicates a relationship between nodes of different degrees. The network matching degree $r$ is defined as:
%\begin{equation}\label{eq:B-i}
%r\!=\!\dfrac{M^{-1} \sum_{m \in G}{j_m k_m}\! -\! [ M^{-1} \sum_{m \in G}{\dfrac{1}{2}(j_m+k_m)} ]^2 }{ M^{-1}\! \sum_{m \in G}{\dfrac{1}{2}(j_m^2+k_m^2)}\! - \![ M^{-1}\! \sum_{m \in G}{\dfrac{1}{2}(j_m+k_m)} ]^2  },
%\end{equation} 
%where $j_m$ and $k_m$ refer to the degrees of the two nodes corresponding to the $m$th edge. $M$ represents the total number of edges. When $r > 0$, the network is positively correlated. Similarly, the network is negatively correlated or uncorrelated when $r < 0$ or $r = 0$, respectively.

\re{In this part, $e_{ij}$ represents the ratio of the number of edges connecting the \re{node $i$ and the node $j$} to all the edges of the network, and express $a_i = \sum_{j \in G}(e_{ij})$, $b_j = \sum_{i \in G}(e_{ij})$. If the network is undirected, then the following formulas are satisfied: $e_{ij}=e_{ji}$, $a_i=b_i$\cite{AssortativityNewman}. Then the {Assortativity coefficient} $r$ is defined as:}
\begin{equation}\label{eq:r}
\re{
r = \dfrac{\sum_{}{e_{ij}} - \sum_{}{a_ib_i}}{ 1-\sum_{}{a_ib_i} }.}
\end{equation}

\subsection{Statistical analysis}

\re{In this study, we use GAT toolbox for graph theoretical analysis~\cite{GAT}.} \re{Specifically, we randomly permute subjects' data between the two groups, for a total of $1000$ times, to generate $1000$ sets of random groups of the same size as the original group size. We then construct $1000$ sets of random connectivity matrices from the 1000 sets of permutation data, by repeating the steps of generating networks. Additionally, 20 null networks are generated using the degree distribution preservation model (\re{two-tail}ed). In the degree distribution preservation model for generating null networks, edges are swapped rather than removed and added back to the network. Therefore, the model preserves the degree distribution of the input network, and each node has the same degree as that of the original network, but with different centralities (such as a different betweenness centrality).} 

\re{We then compare the between-group differences in graph measures (\eg clustering, characteristic path length, small-worldness, global efficiency, etc.) under different density threshold with corresponding difference in randomly-generated graphs.} \re{We take the minimum connectivity density as the minimum density we discussed in this part, and the biological upper limit density (0.500) as the maximum density\cite{2006UpperDensity,GAT}.} \re{The minimum connectivity density is defined as the minimum threshold such that both groups of network are fully-connected\cite{GAT}. And in this case, the minimum connectivity density in the two groups of cortical thickness network and surface area network are 0.1101 and 0.1400, respectively.} % and we consider the network density higher than 0.500 to be not biologically significant. 
The threshold is set by different matrix densities, and the selected density ranges from 0.1101 to 0.5 for cortical thickness network or 0.1400 to 0.5 for surface area network with 0.02 step size. To compare the overall organization of networks, graph theoretical analysis is utilized to extract the following properties from the graph for both FES and HC groups at each link density, including average clustering coefficient $C_{net}$, characteristic path length $L_{net}$, transitivity $T$, global efficiency $E_{global}$, mean local efficiency $E_{local}$, modularity $Q$, mean node betweenness $B_{node}$, mean edge betweenness $B_{edge}$, assortativity coefficient $r$, and small-world properties $\lambda$, $\gamma$, and $\sigma$.
% Calculate the average clustering $C_{net}$, the characteristic path length $L_{net}$, density, the transitivity, the global efficiency $E_{global}$, the mean local efficiency $E_{local}$, the modularity $Q$, the mean node betweenness $B_{node}$, the mean edge betweenness $B_{edge}$, the assortativity $A$, the small-world property $\lambda$, $\gamma$, and $\sigma$ of all 13 threshold points in the range. 

Then we compare all graph network properties between groups with a permutation test. By using random networks to generate 95\% confidence intervals, we analyze the significance level of the differences in various network attributes between the two groups. For the density ranges that show significant differences in the above properties, we examine the contribution of each brain region to those properties \re{by performing the following operations}. \re{For certain properties obtained by averaging the properties of each node in the global analysis (\eg mean local efficiency and mean clustering coefficient), we restore it to 68 node properties to get the contribution of each ROI. \wk{Especially}, for node betweenness, according to the expression of node betweenness in \textbf{Eq.}\eqref{eq:B-node}, we investigate which edge of this node has \wk{contributed} significantly to the change in its node betweenness. This variation \wk{is} also denoted as edge betweenness.} The non-parametric $p$-value test estimation based on \re{the} permutation test is used for group comparison of the brain networks.
% Then compare all parameters between groups after permutation test. Use random networks to generate 95\% confidence intervals and analyze the significance level of the differences in various network attributes between the two groups. For the density ranges that show significant differences in the above properties, we will specifically explore the contribution of 68 brain regions to the significantly different characteristics.

\section{Results}

\subsection{Reduction in FES cortical thickness%in \zc{frontal and parietal lobes}
}

\begin{figure}[!htpb]
    \centering
    \includegraphics[width=1\linewidth]{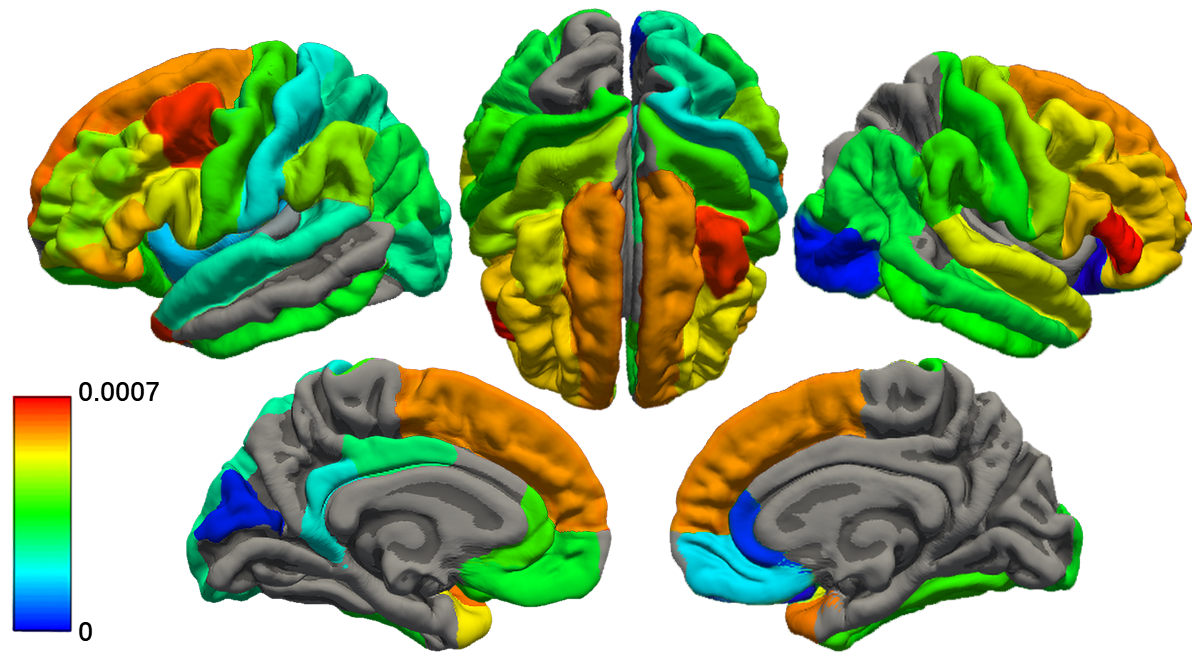}
    \caption{\re{Differences in the cortical thickness of different brain regions between HC and FES groups via permutation tests}. A positive value indicates significantly thicker cortical thickness in the HC group than in the FES group \re{(thresholded at \re{$p<0.05$} after Bonferroni FWER correction). The unit is in millimeters}.}
    \label{fig:thickness}
\end{figure}

\textbf{Figure~\ref{fig:thickness}} shows the \re{ differences in cortical thickness and surface area between FES and HC groups. As we can see,} the brain regions with significant differences in cerebral cortical thickness after FWER correction are mainly concentrated in the frontal and parietal lobes. \re{However, there is no significant brain regions survive after correction for surface area.} \ql{A full list is shown in \textbf{Table}~\ref{tab:ROIs}.}

%\todo{change the figure}

%We found the thickness of the following brain regions in FES group was significantly lower than that in HC group: left and right caudal middle frontal lobe, right fusiform, left and right inferior parietal lobe, left and right inferior temporal artery, left isthmus cingulate gorge, left and right lateral occipital, left and right lateral orbital frontal, left and right medial orbital frontal, right middle temporal, left and right pars opercularis, left and right pars orbital, left and right pars triangularis, left and right postcentral, left posterior cingulate gyrus, left and right rostral anterior cingulate gyrus, left and right rostral middle frontal lobe, left and right superior frontal lobe, left superior parietal lobe, left and right superior temporal artery, left and right supramarginal, left and right temporal pole, left insula.

\begin{table*}
\centering 
\caption{\ql{Statistical analysis on} cortical thickness (in mm) of ROIs. \ql{The brain regions with significant differences between the FES group and HC groups are reported (\re{$p<0.05$} after Bonferroni FWER correction).}}
\small
\begin{tabular}{c|l|c|c|c}
\hline
\textbf{No.}& \textbf{ROI name}&\textbf{FES group}  & \textbf{HC group} &	\textbf{$P$ value}\\ 
\hline
1 & bankssts (left) & 2.6153±0.1605 & 2.6925±0.1408 & 0.0002 \\
2 & caudal middle frontal (left) & 2.5388±0.1763 & 2.6413±0.1559 & \textless{}e-4 \\
3 & fusiform (left) & 2.6988±0.1567 & 2.7727±0.1447 & 0.0002 \\
4 & inferior parietal (left) & 2.4052±0.1239 & 2.4669±0.1289 & 0.0003 \\
5 & lateral orbito frontal (left) & 2.6252±0.1511 & 2.697±0.1333 & 0.0006 \\
6 & medial orbito frontal (left) & 2.4856±0.1601 & 2.5507±0.1407 & 0.0001 \\
7 & middle temporal (left) & 2.9227±0.1493 & 2.9975±0.1461 & 0.0001 \\
8 & pars opercularis (left) & 2.6288±0.1651 & 2.7266±0.1385 & \textless{}e-4 \\
9 & pars orbitalis (left) & 2.6944±0.2268 & 2.8047±0.1895 & 0.0001 \\
10 & pars triangularis (left) & 2.5583±0.1551 & 2.6656±0.1449 & \textless{}e-4 \\
11 & precentral (left) & 2.6052±0.1777 & 2.6815±0.1459 & 0.0006 \\
12 & rostral anterior cingulate (left) & 2.8772±0.1844 & 2.9536±0.1628 & 0.001 \\
13 & rostral middle frontal (left) & 2.3688±0.1421 & 2.4697±0.124 & \textless{}e-4 \\
14 & superior frontal (left) & 2.7596±0.1914 & 2.9075±0.1483 & \textless{}e-4 \\
15 & superior temporal (left) & 2.8507±0.177 & 2.9667±0.1549 & \textless{}e-4 \\
16 & supramarginal (left) & 2.4937±0.1346 & 2.5808±0.124 & \textless{}e-4 \\
17 & insula (left) & 2.8986±0.1752 & 2.9992±0.1481 & \textless{}e-4 \\
18 & caudal middle frontal (right) & 2.5523±0.1746 & 2.6512±0.1507 & \textless{}e-4 \\
19 & fusiform (right) & 2.719±0.153 & 2.7948±0.1524 & 0.0002 \\
20 & inferior parietal (right) & 2.4093±0.1317 & 2.4711±0.1268 & 0.0003 \\
21 & inferior temporal (right) & 2.7522±0.1333 & 2.8221±0.1468 & 0.0001 \\
22 & lateral occipital (right) & 2.0856±0.1087 & 2.1514±0.1191 & 0.0001 \\
23 & medial orbito frontal (right) & 2.4802±0.1465 & 2.5711±0.1432 & \textless{}e-4 \\
24 & pars opercularis (right) & 2.6295±0.1596 & 2.7354±0.1434 & \textless{}e-4 \\
25 & pars orbitalis (right) & 2.671±0.2071 & 2.7746±0.1689 & \textless{}e-4 \\
26 & pars triangularis (right) & 2.5214±0.1386 & 2.6495±0.1419 & \textless{}e-4 \\
27 & postcentral (right) & 2.0568±0.129 & 2.1254±0.1127 & \textless{}e-4 \\
28 & precentral (right) & 2.6054±0.1789 & 2.6956±0.1634 & 0.0001 \\
29 & rostral middle frontal (right) & 2.3671±0.1372 & 2.4721±0.1305 & \textless{}e-4 \\
30 & superior frontal (right) & 2.7614±0.1963 & 2.915±0.1539 & \textless{}e-4 \\
31 & superior temporal (right) & 2.7763±0.1723 & 2.8734±0.1665 & 0.0002 \\
32 & supramarginal (right) & 2.5243±0.1384 & 2.5992±0.1186 & \textless{}e-4 \\
33 & temporal pole (right) & 3.8041±0.2542 & 3.9187±0.2442 & 0.0007 \\
\hline
\end{tabular}
\label{tab:ROIs}
\end{table*}

\subsection{Global analysis}
We then investigate the cortical morphological networks constructed based on the cortical thickness and surface area, respectively. \re{For cortical thickness network under different thresholds (0.1101 $\sim$ 0.500 with a 0.02 step size), we find that the obtained network parameters of the two groups are in the range of $\gamma = C_{net}^{real} / C_{net}^{Rand} > 1 $ and $\lambda = L_{net} ^ {real} / L_{net} ^ {Rand} \sim 1 $.} It proves the small-world network properties of the cortical thickness networks from both groups. Similarly, the surface area networks constructed at each threshold in the range of 0.1400 $\sim$ 0.500 with the step size of 0.02 also result in small-worldness.
% Get the conclusion that the brain function network of FES patients and healthy control group has the nature of small-worldness. 

\ql{We further compare the} differences in brain network \ql{properties} between FES group and HC group at the respective minimum connectivity density of cortical thickness and surface area network. \ql{The results show that} the small-worldness of FES group has \ql{been significantly changed} (\wk{\textbf{Table}~\ref{tab:WholeBrain}}). The clustering coefficient ($C$), transitivity ($T$), average local efficiency ($E_{local}$), and characteristic small-worldness ($\sigma$ and $\gamma$) of cortical thickness network in patients with FES group are significantly smaller than those in HCs. %The transitivity of the surface area network in FES patients is significantly lower than that in the HC control group, 
%\ql{while} the node betweenness and edge betweenness of surface area network are significantly larger in the FES group (\re{$p<0.05$} \ql{after correction in \re{two-tail} permutation test analysis}). 

\begin{table}
\centering 
    \caption{The network parameters and corresponding group-specific topological parameter $p$-values calculated by FES-vs-HC group comparisons via permutation test at the \re{minimum connectivity threshold.} The threshold for the Cortical Thickness and Surface Area are 0.1101 and 0.1400, respectively. Statistical significance $p$-values (\re{$p<0.05$}) are highlighted in the table.} %The positive value means that the parameter of HC group is larger than that of FES group.}
\re{
\small
\begin{tabular}{l|c|c}
\hline
 \textbf{ }&\textbf{Cortical Thickness}  & \textbf{Surface Area} \\ 
 \hline
\textbf{Threshold=}&\textbf{0.1101}&\textbf{0.1400}\\
\hline
$Gamma$&\textbf{0.032}&0.079\\
$Sigma$&\textbf{0.032}&0.078\\
$Lambda$&0.501&0.430\\
$C$&\textbf{0.034}&0.997\\
$L$&0.259&0.769\\
$T$&\textbf{0.027}&\textbf{0.048}\\
$B_{edge}$&0.263&0.114\\
$B_{nodal}$&0.240&0.122\\
$Q$&0.386&0.069\\
$E_{global}$&0.671&0.928\\
$E_{local}$&\textbf{0.033}&0.327\\
$r$&0.321&0.578\\
 \hline
\end{tabular}}
\label{tab:WholeBrain}
\end{table}

\subsection{Regional Analysis}

\begin{figure*}[!htpb]
    \centering
    \includegraphics[width=1\linewidth]{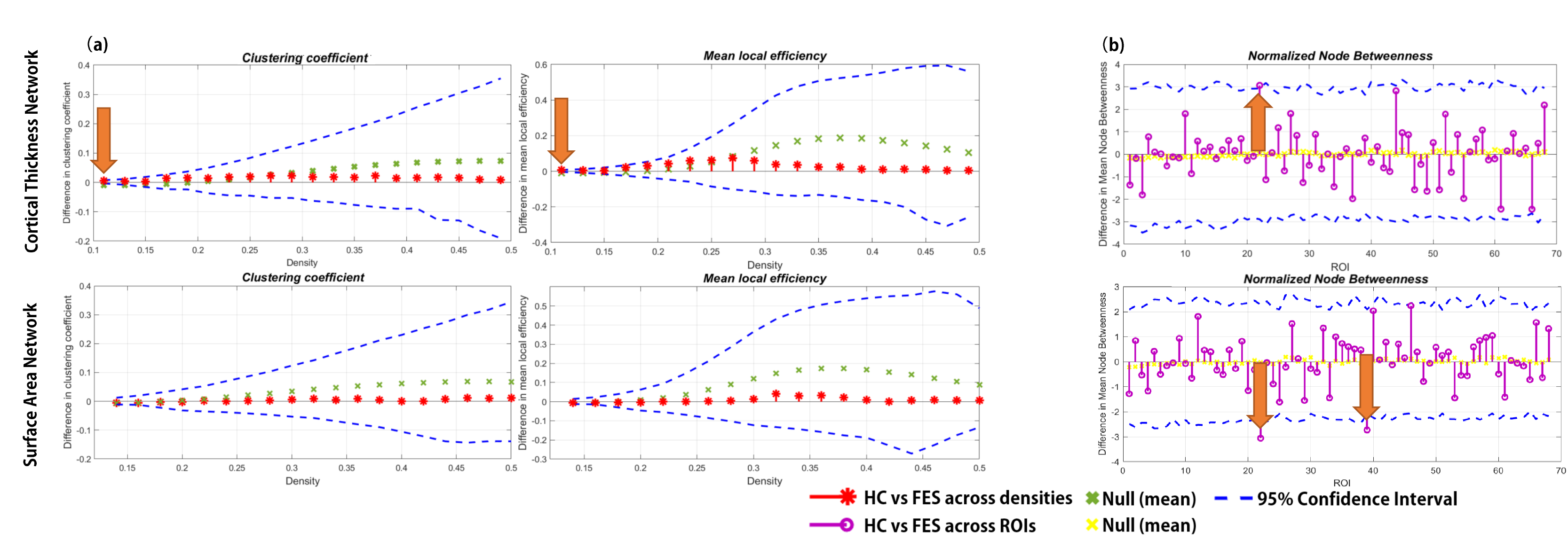}
    \caption{The group differences in certain graphic measurements and 95\% confidence interval (a) for the selected densities, or (b) for 68 brain regions. Within the low-density region, the clustering coefficient and mean local efficiency of the cortical thickness network of FES diminished significantly, \re{highlighted by orange arrows in the figure. The Null lines refer to the mean value of the null hypothesis.}
    }
    \label{fig:crossdensity}
\end{figure*}

%revised
Based on the results of \re{\textbf{Figure} \ref{fig:crossdensity}.(a)}, we find that in both the cortical thickness network and surface area network, the clustering coefficient and mean local efficiency of the HC group are greater than those of the FES group in the low-density interval. As the threshold density increases, when the network density reaches 0.25, the clustering coefficient and local efficiency of the FES group both become greater than those of the HC group, but the difference is not significant. \re{These two parameters only exhibit significant differences at a density of 0.1101. Therefore, we decide to use the minimum density of the Cortical Thickness network (0.1101) for further analysis.}\\
\re{We then further extract the data with the threshold density of 0.1101 and conduct a group-by-group comparison of clustering coefficient and local efficiency (\re{$p < 0.05$} after correction in two-tail permutation test analysis) for all 68 nodes of the network.} 

\begin{figure}[!htpb]
    \centering
    \includegraphics[width=1.0\linewidth]{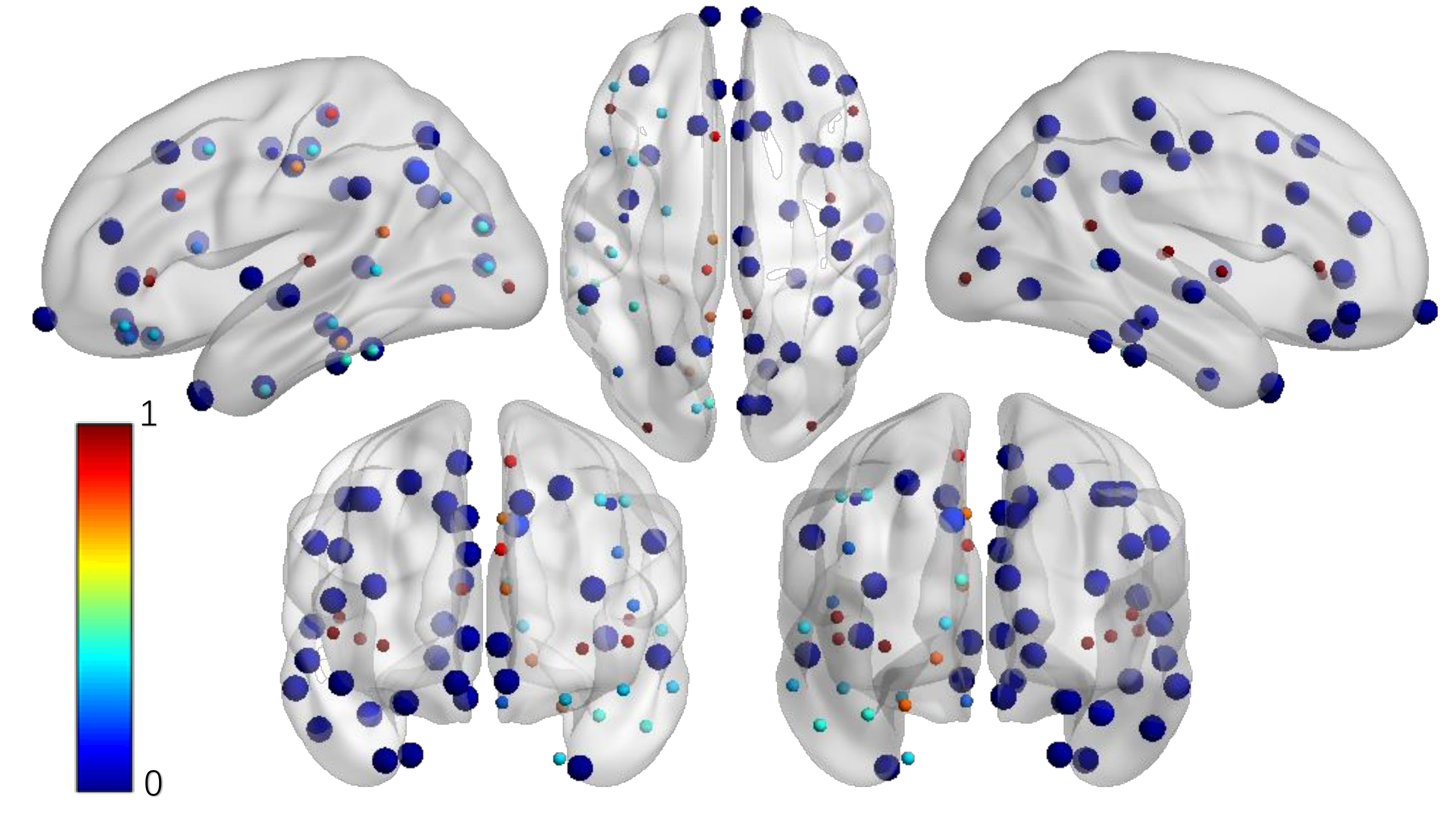}
    \caption{\re{Brain regions with significantly reduced local efficiency}. The color bar reflects the $p$-value. Significant brain regions are colored dark blue and high-lighted with size (\re{$p<0.05$}).}
    \label{fig:ClustLocal}
\end{figure}

Then we compare the nodal clustering coefficients and local efficiency between the two groups.
The full result for regions showing significant differences is shown in \wk{\textbf{Table}~\ref{tab:4Regional}}. %the clustering coefficient and local efficiency of following brain regions increased significantly: left brain, left precuneus, rostral anterior cingulate, rostral middle frontal, superior frontal, superior parietal, superior temporal, supramarginal, frontal pole, temporal pole and insula. For right brain, right bankssts, caudal anterior cingulate, caudal middle frontal, cuneus, entorhinal, fusiform, inferior parietal, inferior temporal, lateral orbitofrontal, lingual, medial orbitofrontal, middle temporal, para hippocampal, paracentral, pars opercularis, pars orbitalis, pericalcarine, postcentral, posterior cingulate, precentral, precuneus, rostral anterior cingulate, rostral middle frontal, superior frontal, superior parietal, superior temporal, supramarginal, frontal pole and temporal pole. 
Most of the brain areas with significant differences are concentrated in the right \ql{hemisphere of the brain}, showing \ql{the dominance of the left hemisphere in the brain of FES patients} (\textbf{Figure} \ref{fig:ClustLocal}).
 
%revised  
\begin{table}
\centering 
\caption{\re{The brain regions show that some have achieved statistical significance in FES-vs-HC group comparisons via $2-tail$ permutation test ($p<0.05$).} And the corresponding local efficiency and clustering coefficient values in two groups with $p$-values calculated at the minimum connectivity density. The threshold density for the cortical thickness network is 0.1101.}
\begin{adjustbox}{width=1\columnwidth}
\begin{tabular}{@{}l|ccc|ccc}
\hline
\multirow{2}{*}{\textbf{brain region (Threshold $=$ 0.1101)}} & \multicolumn{3}{c|}{\textbf{local efficiency}}  & \multicolumn{3}{c}{\textbf{clustering coefficient}} \\ \cline{2-7} & FES & HC & $p$-value  &FES  & HC  & $p$-value\\ 
\hline
precuneus   (left) & 0 & 0 & 0.026 & 0 & 0 & 0.026 \\
rostral anterior cingulate   (left) & 0 & 0 & 0.002 & 0 & 0 & 0.002 \\
rostral middle frontal (left) & 0 & 0 & 0.002 & 0 & 0 & 0.002 \\
superior frontal (left) & 0 & 0 & 0.002 & 0 & 0 & 0.002 \\
superior parietal (left) & 0 & 0 & 0.002 & 0 & 0 & 0.002 \\
superior temporal (left) & 0 & 0 & $<$e-4 & 0 & 0 & $<$e-4 \\
supramarginal (left) & 0 & 0 & $<$e-4 & 0 & 0 & $<$e-4 \\
frontal pole (left) & 0 & 0 & 0.002 & 0 & 0 & 0.002 \\
temporal pole (left) & 0 & 0 & $<$e-4 & 0 & 0 & $<$e-4 \\
insula (left) & 0 & 0 & 0.002 & 0 & 0 & 0.002 \\
bankssts  (right) & 0 & 0 & 0.002 & 0 & 0 & 0.002 \\
caudal anterior cingulate   (right) & 0 & 0 & $<$e-4 & 0 & 0 & $<$e-4 \\
caudal middle frontal (right) & 0 & 0 & 0.004 & 0 & 0 & 0.004 \\
cuneus (right) & 0 & 0 & $<$e-4 & 0 & 0 & $<$e-4 \\
entorhinal (right) & 0 & 0 & 0.004 & 0 & 0 & 0.004 \\
fusiform (right) & 0 & 0 & 0.002 & 0 & 0 & 0.002 \\
inferior parietal (right) & 0 & 0 & $<$e-4 & 0 & 0 & $<$e-4 \\
inferior temporal (right) & 0 & 0 & 0.004 & 0 & 0 & 0.004 \\
lateral orbitofrontal (right) & 24.659 & 0 & 0.002 & 24.659 & 0 & 0.002 \\
lingual (right) & 0 & 0 & 0.004 & 0 & 0 & 0.004 \\
medial orbitofrontal (right) & 0 & 0 & $<$e-4 & 0 & 0 & $<$e-4 \\
middle temporal (right) & 0 & 0 & 0.004 & 0 & 0 & 0.004 \\
parahippocampal (right) & 10.462 & 0 & 0.002 & 10.462 & 0 & 0.002 \\
paracentral (right) & 0 & 0 & 0.002 & 0 & 0 & 0.002 \\
pars opercularis (right) & 0 & 0 & $<$e-4 & 0 & 0 & $<$e-4 \\
pars orbitalis (right) & 0 & 0 & $<$e-4 & 0 & 0 & $<$e-4 \\
pericalcarine (right) & 0 & 0 & $<$e-4 & 0 & 0 & $<$e-4 \\
postcentral (right) & 0 & 0 & 0.002 & 0 & 0 & 0.002 \\
posterior cingulate (right) & 0 & 0 & $<$e-4 & 0 & 0 & $<$e-4 \\
precentral (right) & 0 & 0 & 0.006 & 0 & 0 & 0.006 \\
precuneus (right) & 0 & 0 & $<$e-4 & 0 & 0 & $<$e-4 \\
rostral anterior cingulate   (right) & 0 & 0 & 0.004 & 0 & 0 & 0.004 \\
rostral middle frontal   (right) & 0 & 0 & 0.002 & 0 & 0 & 0.002 \\
superior frontal (right) & 0 & 0 & $<$e-4 & 0 & 0 & $<$e-4 \\
superior parietal (right) & 0 & 0 & 0.004 & 0 & 0 & 0.004 \\
superior temporal (right) & 0 & 0 & 0.002 & 0 & 0 & 0.002 \\
supramarginal (right) & 0 & 0 & 0.006 & 0 & 0 & 0.006 \\
frontal pole (right) & 32.879 & 0 & $<$e-4 & 32.879 & 0 & $<$e-4 \\
temporal pole (right) & 0 & 0 & 0.002 & 0 & 0 & 0.002\\
\hline
\end{tabular}
\end{adjustbox}
\label{tab:4Regional}
\end{table}

\subsection{Nodal Analysis}
%revised
\begin{table}[b]
\centering
\caption{\re{Regions with statistically significant contributions in the node betweenness contribution to PCC (left)}. (\re{$p<0.05$} via permutation test)} %The threshold density is set at the minimum connection density (0.1101). }
\begin{tabular}{l|ccc}
\hline
\multicolumn{4}{c}{Node betweenness to left PCC} \\
\hline
ROI & FES & HC & $p$-value \\
\hline 
superior temporal (left) & 42.02 & 0 & 0.007 \\
supramarginal (left) & 0 & 0 & 0.05 \\
parsorbitalis (right) & 0 & 0 & 0.023 \\
\hline
\end{tabular}
\label{tab:NodalCTT}
\end{table}

%revised
%\begin{sidewaystable}
\begin{table*}
\centering 
\caption{The regions with statistically significant difference (\re{$p<0.05$ via} permutation test) on topological parameter node betweenness between brain regions and left PCC or right entorhinal as well as the corresponding $p$-values in FES-vs-HC group comparisons. The threshold density is set at the minimum \re{connectivity} density (0.1400).}
\scalebox{1.05}{
\begin{tabular}{@{}llll|llll@{}}
\toprule
\multicolumn{8}{c}{\cellcolor[HTML]{FFFFFF}{\color[HTML]{333333} \textbf{Node betweenness for surface area at density=0.1400}}} \\ \midrule
\multicolumn{4}{l}{\cellcolor[HTML]{FFFFFF}posterior cingulate (left)} & \multicolumn{4}{l}{\cellcolor[HTML]{FFFFFF}entorhinal (right)} \\
\hline
\textbf{ROI} & \textbf{FES} & \textbf{HC} & \textbf{$p$-value} & \textbf{ROI} & \textbf{FES} & \textbf{HC} & \textbf{$p$-value} \\
\hline
cuneus (left) & 8.65 & 0.00 & 0.018 & bankssts (left) & 10.92 & 0.00 & 0.006 \\
parsorbitalis (left) & 9.28 & 0.00 & 0.007 & lingual (left) & 0.00 & 0.00 & 0.006 \\
postcentral (left) & 0.00 & 0.00 & 0.018 & paracentral (left) & 8.46 & 0.00 & 0.020 \\
precentral (left) & 0.00 & 0.00 & 0.023 & pericalcarine (left) & 7.87 & 0.00 & 0.024 \\
precuneus (left) & 0.00 & 0.00 & 0.007 & postcentral (left) & 11.14 & 0.00 & 0.004 \\
rostral anterior cingulate (left) & 10.82 & 0.00 & 0.006 & precentral (left) & 12.46 & 0.00 & 0.003 \\
rostral middle frontal (left) & 0.00 & 0.00 & 0.024 & precuneus (left) & 9.84 & 0.00 & 0.007 \\
superior parietal (left) & 9.86 & 0.00 & 0.009 & rostral anterior cingulate (left) & 6.51 & 0.00 & 0.017 \\
superior temporal (left) & 0.00 & 0.00 & 0.023 & caudal middle frontal(right) & 0.00 & 0.00 & 0.011 \\
supramarginal (left) & 10.62 & 0.00 & 0.004 & fusiform (right) & 0.00 & 0.00 & 0.022 \\
frontal pole (left) & 8.32 & 0.00 & 0.015 & inferior parietal (right) & 0.00 & 0.00 & 0.022 \\
temporal pole (left) & 12.63 & 0.00 & 0.004 & lateral orbitofrontal (right) & 0.00 & 0.00 & 0.014 \\
isthmus cingulate (right) & 9.33 & 0.00 & 0.018 & medial orbitofrontal (right) & 11.67 & 0.00 & 0.006 \\
medial orbitofrontal (right) & 0.00 & 0.00 & 0.017 & middle temporal (right) & 7.22 & 0.00 & 0.006 \\
middle temporal (right) & 9.81 & 0.00 & 0.010 & parsopercularis (right) & 6.88 & 0.00 & 0.017 \\
parsopercularis (right) & 6.52 & 0.00 & 0.021 & parstriangularis (right) & 0.00 & 0.00 & 0.020 \\
parsorbitalis (right) & 0.00 & 0.00 & 0.010 & precentral (right) & 0.00 & 0.00 & 0.024 \\
precentral (right) & 9.31 & 0.00 & 0.014 & superior frontal (right) & 0.00 & 0.00 & 0.018 \\
precuneus (right) & 11.07 & 0.00 & 0.015 & supramarginal (right) & 9.22 & 0.00 & 0.012 \\
rostral anterior cingulate (right) & 14.40 & 0.00 & 0.002 & transverse temporal (right) & 0.00 & 0.00 & 0.023 \\
rostral middle frontal (right) & 7.51 & 0.00 & 0.021\\
superior temporal (right) & 0.00 & 0.00 & 0.018\\
supramarginal (right) & 0.00 & 0.00 & 0.050\\
\bottomrule
\end{tabular}
}
\label{tab:NodalSA}
\end{table*}
%\end{sidewaystable}

\begin{figure}[!htpb]
    \centering
    \includegraphics[width=1.0\linewidth]{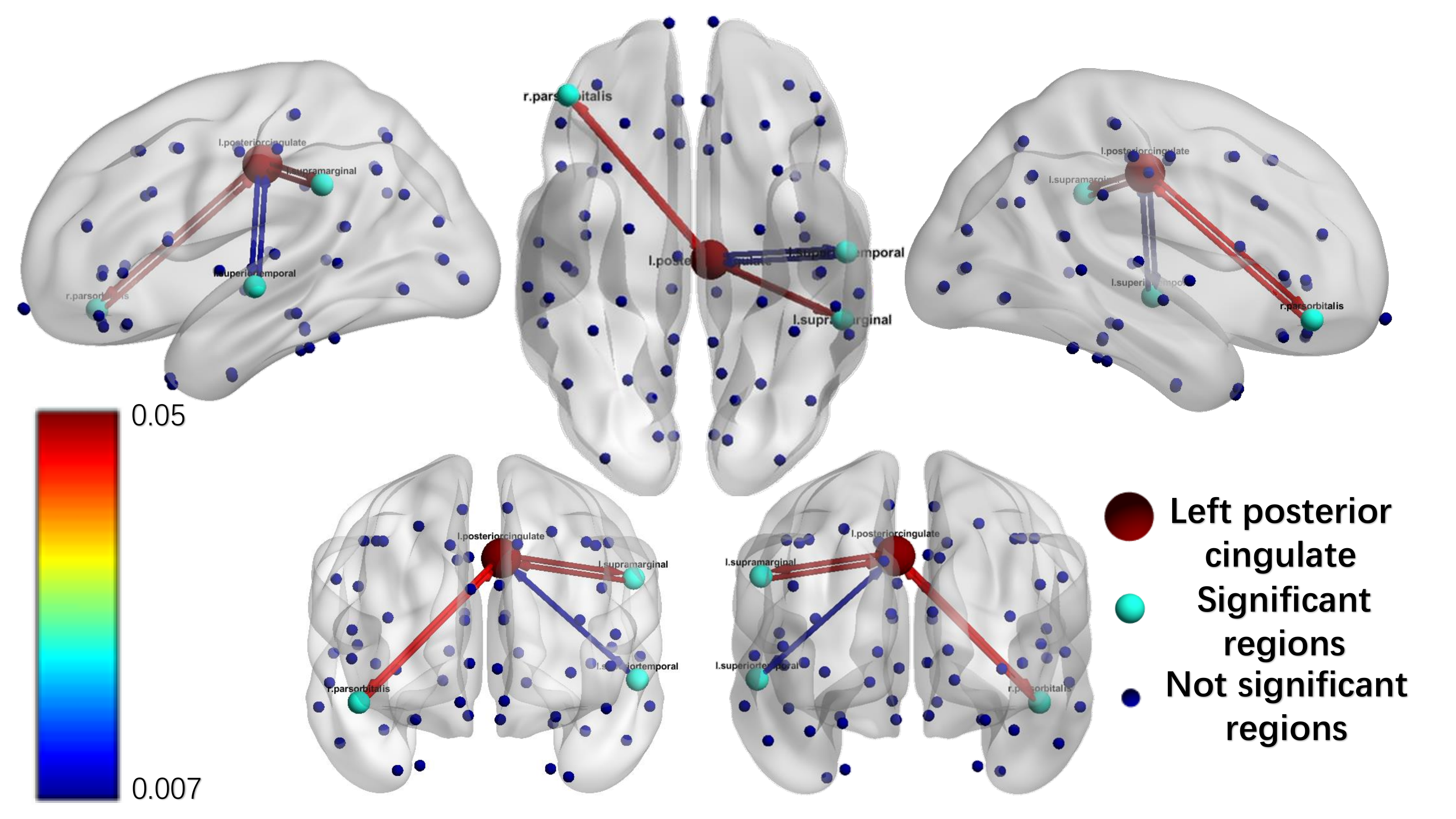}
    \caption{The most prominent brain connections and corresponding brain regions (light blue) that affecting \re{the} edge betweenness of the left PCC nodes (red). The color bar corresponds to the color of the connecting edges in the figure, reflecting $p$-values.}
    \label{fig:CT PCC}
\end{figure}

%revised
We then study the regions that show significant differences in node betweenness (\textbf{Figure} \ref{fig:crossdensity}). For Cortical Thickness network, the $22$nd brain region (left PCC) reaches a significant difference. We calculate the edge betweenness from each node to the PCC (left) at threshold density = 0.1101, and derive regions with significant differences \re{(two-tail, $p < 0.05$)}. The results show that for left PCC, the following three nodes show significant differences in edge betweenness (\textbf{Table} \ref{tab:NodalCTT}): left superior temporal, left supramarginal, and right pars orbitalis. The distribution in the brain is shown in \textbf{Figure} \ref{fig:CT PCC}.

Similarly, for the surface area network, both the left PCC and the right entorhinal show significant differences. We do the same thing on the surface area network, calculating the edge betweenness of each node to left PCC or right entorhinal, respectively. The full results are listed \wk{in \textbf{Table}}~\ref{tab:NodalSA}. 
Firstly, according to the $p$-value of edge betweenness to left PCC, the node betweenness of the FES group from the following brain regions to left PCC have significantly increased: left cuneus, pars orbitalis, postcentral, precentral, precuneus, rostral anterior cingulate, rostral middle frontal, superior parietal, superior temporal, supramarginal, frontal pole, temporal pole, isthmus cingulate, right medial orbitofrontal, middle temporal, pars opercularis, pars orbitalis, precentral, precuneus, rostral anterior cingulate, rostral middle frontal, superior temporal, and supramarginal. Among them, left superior temporal and right pars orbitalis are important nodes that show significant differences both in cortical thickness and surface area networks.
Secondly, for the right entorhinal, our results show the following regions significantly \re{increased} on edge betweenness: left bankssts, lingual, paracentral, pericalcarine, postcentral, precentral, precuneus, rostral anterior cingulate, right caudal middle frontal, fusiform, inferior parietal, lateral orbitofrontal, medial orbitofrontal, middle temporal, pars opercularis, pars triangularis, precentral, superior frontal, supramarginal, and transverse temporal.

\subsection{\ree{Replication using Brainnetome atlas}}

\ree{We replicate the cortical thickness network construction using Brainnetome atlas. We find the minimum connectivity density of FES cortical thickness network is 0.79 while it is 0.32 for the HC group.}

\begin{figure*}[!htpb]
    \centering
    \includegraphics[width=0.8\linewidth]{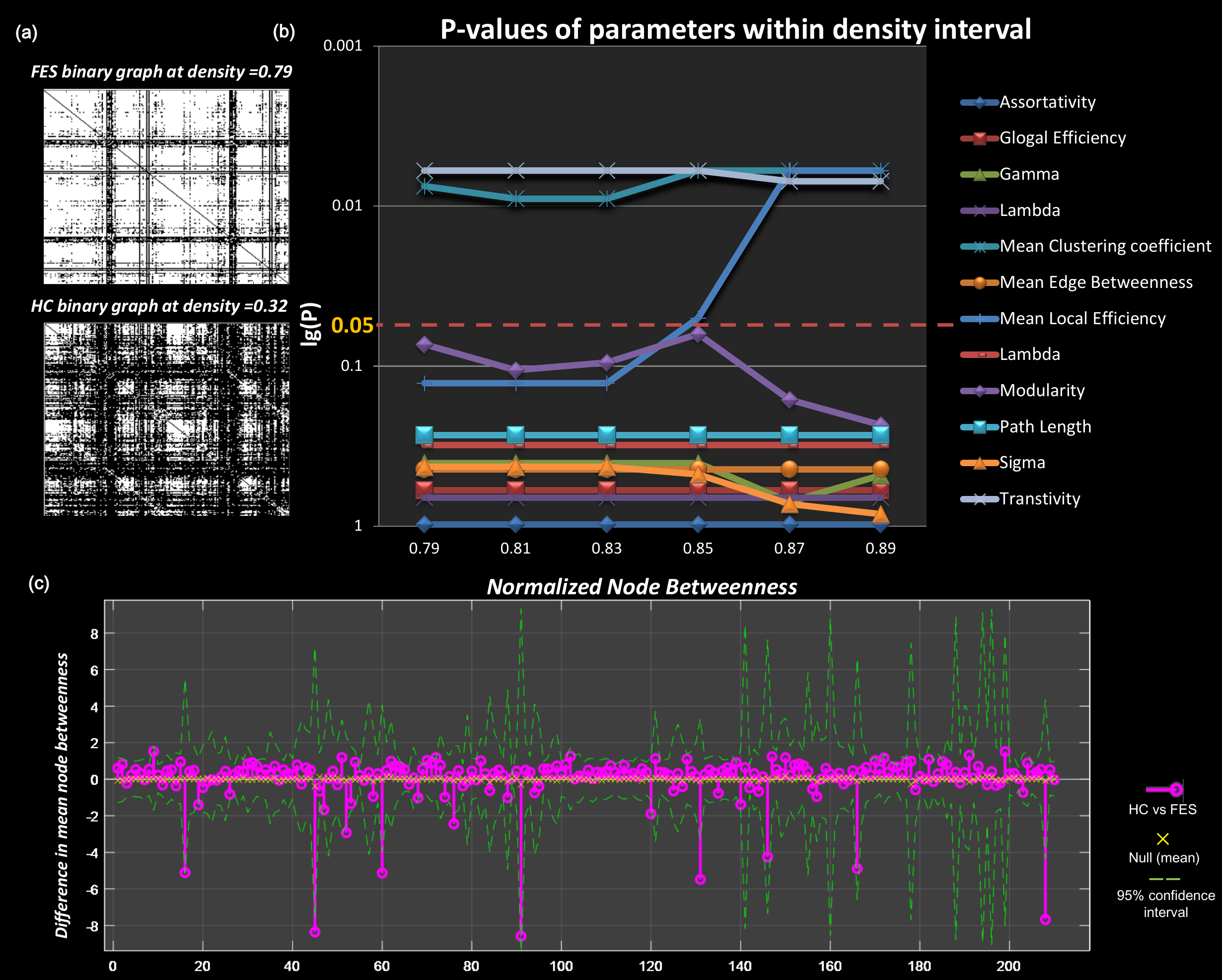}
    \caption{\ree{Replication using Brainnetome atlas. (a) Binary networks of both groups at their corresponding minimum connectivity densities. (b) The p-values for graph theoretical parameters within a specific density interval, suggesting that transitivity, mean clustering coefficient and mean local efficiency with densities $\geq$ 0.85 show significant differences (p $\leq$ 0.05, after FDR correction). (c) Normalized node betweenness of all 210 brain regions (before FDR correction).} }
    \label{fig:BNAcross}
\end{figure*}

\ree{We then replicate graph theoretical analysis on the cortical thickness network obtained with Brainnetome atlas. We calculate graph theoretical parameters across density interval (0.79 $\sim$ 0.89). The results are shown in Figure~\ref{fig:BNAcross}.(b), which are generally but not exactly consistent with the results with the Desikan-Killiany atlas. Specifically, the mean clustering coefficient, transitivity, and mean local efficiency with densities $\geq$ 0.85 show significant differences (p $\leq$ 0.05, after FDR correction). We further examine the abnormal node betweenness of the cortical thickness network. Figure~\ref{fig:BNAcross}.(c) shows several brain regions show significant decreases in node betweenness. However, none of them survive after FDR correction for a more refined atlas largely increases the number of comparisons.}

\section{Discussion}

The pathological mechanism of schizophrenia across the human lifespan has long been an open question. \wk{Linking} human brain macroscale network topology to cognitive function and clinical disorder is a major method for studying the pathology of psychosis\cite{BrainNetwork}. \ql{A number of studies have reported the changes in the CMN under certain pathological conditions\wk{, such as the cortical thickness and the surface area of the cerebral cortex, which are characteristic of brain development and aging}~\cite{Schinack2015,2Tang2021}.} 
%For example, the cortical thickness and the surface area of the cerebral cortex where characterize the development and aging of brain.
\wk{For more specific examples,} posterior frontal and superior parietal lobes are associated with intelligence~\cite{BAJAJ201836,Schinack2015}, and ADHD and autism show an overall decrease in cortical thickness covariance~\cite{CMNonADHD_autism}. However, the abnormality related to FES is still not clear.

In this study, we investigate the CMN in FES in terms of cortical thickness and surface area. We find the cortical thickness network and the surface area network satisfy the small-world property, and at certain levels of network sparsity, the clustering coefficient and local efficiency of the cortical thickness network and the surface area network in the FES group are significantly lower than those in the HC group, indicating FES-related local connectivity and small-world abnormalities. Abnormal regions mainly involve the frontal, parietal and temporal lobes, and these regions are closely related to brain cognitive function. In further nodal analysis, we find that there are significant differences in the node betweenness in the PCC between the two analysis networks, and the node betweenness of the entorhinal in the surface area network is also significantly decreased.

%This organizational model that combines the specialization of brain function and the integration of brain function allows the brain to achieve efficient and complex functional activities\cite{Sophie2007013}. 
%The changes in brain structure and organization have crucial effects on the information process and exchange across brain regions, especially on the complex and higher-order neural activities. 
%\wk{Researches have found that} under certain pathological conditions, \wk{some} structural changes will occur in the cerebral cortex. For example, the thickness and the surface area of the cerebral cortex characterize\wk{, to a certain extent,} the development and aging of brain tissue and are \wk{associated with} diseases such as neurasthenia\cite{Schinack2015}. These changes in brain structure and organization \wk{have} a crucial impact in the exchange of information \wk{between} brain regions, especially the complex and \wk{higher-order} activities of the brain.

\subsection{The whole-brain attribute affects small-worldness}
The group comparisons of cortical thickness between FES and HC in 68 cortical regions showed that \ql{the abnormality in FES covers a large area of the brain} (\textbf{Figure} \ref{fig:thickness}). Among them, the reduction of anterior cingulate, middle and inferior frontal, insular, and middle and superior temporal regions are \wk{consistently associated with cognitive impairment in schizophrenia, which is} in line with previous studies\cite{WANG2021179,SUI2015794}. 
Our results show that the abnormal regions \wk{are mainly clustered in} the prefrontal and temporal lobes, \ql{regions considered to be developing until substantial maturation during adolescence\cite{CABALLERO20164}. Interestingly,} adolescence happens to be a period of increased susceptibility to schizophrenia, which might be related to abnormal developments of prefrontal and temporal lobes in FES~\cite{BENOIT2022}. Our findings are consistent with previous findings that \wk{prefrontal cortex thinning} is highly associated with the occurrence of schizophrenia, even in the mild or absence of cognitive impairment\cite{HANFORD2019129}.

\ql{We further conduct the graph theoretical analysis on the constructed cortical morphological networks. We find} \wk{all brain networks in FES and HC groups have the small-world property (\textbf{Table}~\ref{tab:WholeBrain}), \re{and both the cortical thickness network and the surface area network are in the range of the minimum connectivity density $\leq \text{threshold} \leq 0.5$. The minimum densities \wk{of} the cortical thickness and surface area networks are 0.1101 and 0.1400, respectively.}}
Another small-world characteristic value $\gamma$ reflects the local information processing efficiency of the network. $\gamma$ of the brain network of FES patients is significantly smaller than that of the healthy control, \wk{which} means the less network's ability to process information locally. The small-word characteristic value $\lambda$ mainly reflects the information transmission and integration efficiency of the network. The reduction of $\lambda$ indicates that the information processing efficiency of these brain regions is decreased. 
\ql{Our findings are in line with previous studies}, reporting that \wk{small world properties in schizophrenia patients} are changed at whole brain attribute analysis~\cite{JHUNG201335}.

%revised
\ql{We also find that} connections between brain regions in patients with FES are decreased. The clustering coefficient and average local efficiency of cortical morphological networks in FES patients are significantly lower than those in HC group (\textbf{Table} \ref{tab:WholeBrain}). The clustering coefficient reflects the \ql{ability of the network to process local information:} the lower the clustering coefficient, the weaker the \ql{network's} ability to process local information\cite{WattsStrogatz}. \ql{The change of clustering coefficient in patients with FES is still debating.} The lower clustering coefficient of FES group in our study is in line with most of the previous studies\cite{Small-worldNon,Lynall2010022,liu2008disrupted}, however, a few studies got the opposite finding. For instance, \wk{in} Zhou's research, early-onset schizophrenia (EOS) patients showed significantly increased clustering and local efficiency instead of decrease\cite{ZHOU2021027}. This may be a different characteristic between EOS patients and FES patients. Although the characteristics are different, we all draw the conclusion of the patients sacrificed the overall efficiency of the brain network to a certain extent, suggesting that there is a lack of long-term information exchange in different brain regions of schizophrenic patients. This indicates that brain network information transmission ability, which are believed as the basis of cognitive processing, are impaired\cite{Bassett200907}. %Therefore, the impaired information transmission ability and information integration ability can infer that the cognitive function of patients with FES may be impaired.

\subsection{Local brain analysis}
%revised
\wk{In our localize brain analysis, abnormalities in most areas of the cortex are observed} in \textbf{Figure} \ref{fig:ClustLocal}, which are mainly concentrated in the clustering coefficient and local efficiency of the right brain area, thus showing the dominance of left hemisphere in the brain of FES patients. The results of our study (\textbf{Table} \ref{tab:4Regional}) \wk{show} the decrease of node clustering coefficient and local efficiency are mainly in temporal lobe and frontal lobe region.
%revised
Among them, the superior temporal lobe (\eg left superior temporal and right superior temporal) have \wk{a} reduced ability to process local information. It further shows that the information exchange between \wk{superior temporal} and other brain regions of the whole brain slows down, and the integration efficiency decreases. \wk{The main language perception zone, superior temporal lobe, is involved in auditory information processing, language information processing and expression, and thinking\cite{Wei2012017}.} The damage of the superior temporal can cause positive symptoms such as auditory hallucinations or delusions in patients with schizophrenia\cite{Woodward018}.

%revised
\wk{It is noteworthy that we find} the clustering coefficient and local efficiency of the nodes of the right brain frontal lobe (\eg right pars opercularis and pars triangularis) and superior parietal are also reduced. So far, many studies support that those schizophrenic patients generally have obvious memory disorders, especially short-term memory disorders like working memory disorders\cite{Zhao1998019,KANG2018210}. The related regions including the lateral prefrontal cortex and the posterior parietal cortex is about spatial working memory\cite{YAO2021187}, and the activation of the left ventral side of the left frontal lobe and the lower-left posterior parietal lobe is about speech sexual working memory\cite{Liu2004020}. The clustering coefficients and local efficiency abnormalities of these brain regions in our study reflect the defects of information retention and information execution control process in FES patients.

%revised
Furthermore, the reduction in the node clustering coefficient and local efficiency of the prefrontal lobe (\eg posterior frontal gyrus and orbitofrontal cortex) and frontal pole also has clinical meaning. It is currently believed that the prefrontal lobe is connected with the functions of language ability, fine autonomous movement, and advanced cognitive functions such as judgment, decision-making, and information extraction.\cite{HOLMES1989118}. And frontal pole is usually related to cognitive insight\cite{BENOIT2022}. Therefore, the above result may suggest a defect in the cognitive function of FES patients.

Significant decreases in node clustering coefficients are also found in the occipital lobe and the limbic system (\eg left PCC). Beside CMN, previous studies on FCN of schizophrenia patients have also found abnormalities in some topological attributes at occipital lobe and cingulate cortex\cite{Gong55}. The decrease of clustering coefficient in these brain regions is usually accompanied by the change of small-world parameters and the decrease of connection strength\cite{Lynall2010022}. This indicates FES patients \wk{may} have damaged overall cortical-liminal system neural connection circuits. Once the speed of nerve signal transmission between brain regions and the effectiveness of signal transmission is reduced, which ultimately leads to the destruction of brain morphometry networks with relatively unique local and overall organizational structures.

\re{This study further confirms} the hypothesis of the neuropathology of abnormal structural connections in the brain of schizophrenia. The connecting fibers of each brain interval of schizophrenia are damaged, weakening the brain's ability to integrate external input stimulus information, making further related information processing disordered, leading to a series of mental symptoms\cite{HARRISON2008421}.

%\subsection{Nodal connection analysis reveals changes in the importance of PCC nodes in FES brain}
\subsection{Nodal connection analysis}
%revised
According to our results (\textbf{Figure}~\ref{fig:crossdensity}), the PCC shows \wk{a} significant difference in node \wk{betweenness} both in cortical thickness network and surface area network, indicating PCC is a core node in the changes of cortical morphological network in FES patients.

%revised
The cingulate gyrus is part of the marginal lobe, and the PCC is connect to the combined temporal lobe, middle temporal lobe, orbitofrontal cortex and medial occipital lobe. In the node betweenness analysis of cortical thickness and surface area network, the node betweenness from the left PCC to other brain regions decreased significantly in cortical thickness network, while it increased significantly in surface area network. We found the node betweenness of PCC and left superior temporal, left supramarginal, and right pars orbitalis increased (\textbf{Table} \ref{tab:NodalCTT}), which means the role and influence across the network of these connections increased. Previous studies have shown that the functional connection between PCC and subfrontal area is related to the severity of positive and negative symptoms in patients with schizophrenia\cite{LIANG2020026}. Also, the PCC has been shown to be related to visuospatial behavior involving retroocular neuronal response and spatial memory and involved in the retrieval of previously learned data\cite{HAZNEDAR2004025}. Our findings are consistent with these studies, \ql{implying that the cortical connectivity in the PCC} in FES patients may \ql{be} a hallmark change and may be \ql{responsible for the patients'} visual deficits and memory deficits.

%revised
For surface area network, besides PCC, the characteristic change is that the node betweenness of most brain regions to right entorhinal cortex has changed, seen in \textbf{Figure} \ref{fig:crossdensity} and \textbf{Table} \ref{tab:NodalSA}. Early \re{disruption} of its structure may lead to neuropsychological changes in \ql{latter} development and adulthood. Previous studies on the entorhinal cortex in schizophrenia were mostly based on physiological and morphological aspects, such as aberrant invaginations of the surface, disruption of cortical layers, heterotopic displacement of neurons, and paucity of neurons in superficial layers\cite{archpsyc1991028}. Our study \ql{highlights} the importance of entorhinal cortical connectivity in patients with FES from the perspective of \ql{brain} network. The entorhinal cortex is the key of the nervous system that mediates the interaction between cortex and hippocampus. The results indicating several connections together caused the property change of the node betweenness of the entorhinal cortex.

\subsection{Brainnetome atlas analysis}
\ree{In previous studies, it has been suggested that brain networks greater than 50\% being connected are unlikely to be biological networks because a dense network will be increasingly random (small-world index $\sigma \leq$ 1.2).  ~\cite{2006UpperDensity,GAT,ZHOU2021027}. Concerning the results of the calculated minimum density in Figure \ref{fig:BNAcross}.(a), it indicates that the FES network might have different small-world properties such as clustering coefficient and characteristic path length ~\cite{Small-worldNon}. This is also confirmed by the results of Figure ~\ref{fig:BNAcross}.(b). These preliminary results of Brainnetome atla are generally consistent with the results from atlas having 68 brain regions, which to some extent confirms the reliability of our original analyses. We recommend future researchers to also explore other brain atlases such as the AAL atlas to further verify these properties. A more refined analysis will be one of our future research plans.}

\subsection{Significance of research}
%revised
This study has great significance for the diagnosis and treatment of schizophrenia patients. \wk{Until now, diagnoses are generally not made until psychotic symptoms or schizophrenia symptoms are evident in high-risk mental state populations (ARMS).}
However, when the symptoms of psychosis first appeared, the neurobiological process associated with schizophrenia may \wk{have been ongoing for years}. And the above discussion of the findings supports the existence of characteristic cognitive and certain brain dysfunction in patients with schizophrenia. Therefore, by incorporating cognitive decline (judgment based on brain functional structure network) into the diagnostic basis of schizophrenia, and taking improvement of cognitive function as the main direction of treatment research, early intervention can be carried out before patients' brain function deteriorates.

%revised
This study also provides a direction worthy of further exploration for future research on schizophrenia. \re{As early as the concept of schizophrenia was proposed by the psychiatrist Bleuler of the University of Zurich Switzerland in 1911\cite{bleuler1911dementia}, it was believed that the cause of schizophrenia is not the dysfunction of a single specific brain region, but is caused by complex and extensive brain region abnormalities.} \re{Besides our research, recent in-depth researches on the topological structure of brain networks are expected to progressively validate and deepen this hypothesis.} It also provides a valuable way for the neurobiological pathology of schizophrenia. In the future, research in the following directions can be used to continue to explore schizophrenia: exploring the relationship between the severity of FES patients (\eg PANSS score) and the morphological connectivity network properties (\eg small-worldness); combining and optimizing machine learning to conduct large sample and long-term comparison of different subtypes of schizophrenia and other mental diseases on the basis of existing researches~\cite{4YUAN2022109441,2Tang2021}; following up the study of brain function connections at different developmental stages of schizophrenia. 
%\ree{One thing that is worthy of being pointed out is that it would significantly enhance the impact of this work by employing a more refined atlas to characterize more detailed abnormality patterns of the human brain's network topology and to identify the robustness of this work's observations. A good candidate is the Brainnetome atlas \cite{JIANG2013263}, the analyses on which will also be one of our future research plans.} 
Further development also lies in the emphasis on interdisciplinary research, and in-depth integration with other fields, especially gene research and drug therapy. It is expected to analyze the relationship among schizophrenia psychiatric symptoms, brain function imaging abnormalities and neurobiological basis.

%revised
\subsection{Limitations}
However, \wk{our} study still has limitations. First, the sample size is relatively small, and the results reported in this study have to be replicated with larger datasets in the future. Second, there are commonalities between the mechanisms and symptoms of schizophrenic patients and some putative causal factors shared by the diagnosis of other diseases, making it difficult to obtain specific treatments for schizophrenia. Therefore, it may be difficult to improve the intervention measures for schizophrenia targeted at a single mechanism and characteristics, since it was proposed to be more suitable for being regarded as a syndrome instead of a disease\cite{CANNON202212}. Besides, our samples are predominantly Asian, which may limit the accuracy of the results.

\section*{Acknowledgements}
The authors gratefully acknowledge the anonymous reviewers for their insightful comments. %This study is supported by the National Natural Science Foundation of China (62071210, 62001205), the Shenzhen Science and Technology Program (RCYX20210609103056042), the Shenzhen Basic Research Program (JCYJ20200925153847004, JCYJ20190809120205578), the Guangdong Natural Science Foundation Joint Fund (2019A1515111038), the Shenzhen Science and Technology Innovation Committee (20200925155957004, KCXFZ2020122117340001), the Shenzhen Key Laboratory of Smart Healthcare Engineering (ZDSYS20200811144003009).

\bibliographystyle{IEEEtran}
%\bibliography{reference.bib}

\begin{thebibliography}{10}
\providecommand{\url}[1]{#1}
\csname url@samestyle\endcsname
\providecommand{\newblock}{\relax}
\providecommand{\bibinfo}[2]{#2}
\providecommand{\BIBentrySTDinterwordspacing}{\spaceskip=0pt\relax}
\providecommand{\BIBentryALTinterwordstretchfactor}{4}
\providecommand{\BIBentryALTinterwordspacing}{\spaceskip=\fontdimen2\font plus
\BIBentryALTinterwordstretchfactor\fontdimen3\font minus
  \fontdimen4\font\relax}
\providecommand{\BIBforeignlanguage}[2]{{%
\expandafter\ifx\csname l@#1\endcsname\relax
\typeout{** WARNING: IEEEtran.bst: No hyphenation pattern has been}%
\typeout{** loaded for the language `#1'. Using the pattern for}%
\typeout{** the default language instead.}%
\else
\language=\csname l@#1\endcsname
\fi
#2}}
\providecommand{\BIBdecl}{\relax}
\BIBdecl

\bibitem{wiersma1998natural}
D.~Wiersma, F.~J. Nienhuis, C.~J. Slooff, and R.~Giel, ``Natural course of
  schizophrenic disorders: a 15-year followup of a dutch incidence cohort,''
  \emph{Schizophrenia bulletin}, vol.~24, no.~1, pp. 75--85, 1998.

\bibitem{larsen199601}
T.~K. Larsen, T.~H. McGlashan, and L.~C. Moe, ``First-episode schizophrenia: I.
  early course parameters,'' \emph{Schizophrenia bulletin}, vol.~22, no.~2, pp.
  241--256, 1996.

\bibitem{weiden2007understanding}
P.~J. Weiden, P.~F. Buckley, and M.~Grody, ``Understanding and treating
  “first-episode” schizophrenia,'' \emph{Psychiatric Clinics of North
  America}, vol.~30, no.~3, pp. 481--510, 2007.

\bibitem{murray1997global}
C.~J. Murray and A.~D. Lopez, ``Global mortality, disability, and the
  contribution of risk factors: Global burden of disease study,'' \emph{The
  lancet}, vol. 349, no. 9063, pp. 1436--1442, 1997.

\bibitem{MRI}
P.~Lauterbur, ``Image formation by induced local interactions: Examples
  employing nuclear magnetic resonance,'' \emph{Clinical Orthopaedics and
  Related Research}, vol. 244, 07 1989.

\bibitem{Neuroimaging}
\BIBentryALTinterwordspacing
S.~Bunge and I.~Kahn, ``Cognition: An overview of neuroimaging techniques,'' in
  \emph{Encyclopedia of Neuroscience}, L.~R. Squire, Ed., 2009, pp. 1063--1067.
  [Online]. Available:
  \url{https://www.sciencedirect.com/science/article/pii/B9780080450469002989}
\BIBentrySTDinterwordspacing

\bibitem{CT}
B.~Cui, ``analysis of medical imaging technology ct,'' \emph{world's latest
  medical information digest}, vol.~15, no. 072, pp. 111--112, 2015.

\bibitem{PET}
G.~B. Saha, \emph{Basics of PET imaging: physics, chemistry, and
  regulations}.\hskip 1em plus 0.5em minus 0.4em\relax Springer, 2015.

\bibitem{Tang2020}
\BIBentryALTinterwordspacing
X.~Tang, G.~Lyu, M.~Chen, W.~Huang, and Y.~Lin, ``Amygdalar and hippocampal
  morphometry abnormalities in first-episode schizophrenia using
  deformation-based shape analysis,'' \emph{Frontiers in Psychiatry}, vol.~11,
  2020. [Online]. Available:
  \url{https://www.frontiersin.org/article/10.3389/fpsyt.2020.00677}
\BIBentrySTDinterwordspacing

\bibitem{1Tang2021}
W.~Huang, M.~Chen, G.~Lyu, and X.~Tang, ``A deformation-based shape study of
  the corpus callosum in first episode schizophrenia,'' \emph{Frontiers in
  Psychiatry}, vol.~12, 2021.

\bibitem{ADDINGTON199850}
\BIBentryALTinterwordspacing
J.~Addington and D.~Addington, ``Neurocognitive functioning in first episode
  schizophrenia,'' \emph{Schizophrenia Research}, vol.~29, no.~1, pp. 50--51,
  1998. [Online]. Available:
  \url{https://www.sciencedirect.com/science/article/pii/S0920996497884186}
\BIBentrySTDinterwordspacing

\bibitem{Sporns2004015}
O.~Sporns, D.~Chialvo, M.~Kaiser, and C.~Hilgetag, ``Organization, development
  and function of complex brain networks,'' \emph{Trends in cognitive
  sciences}, vol.~8, pp. 418--25, 10 2004.

\bibitem{sporns2016networks}
\BIBentryALTinterwordspacing
O.~Sporns, \emph{{Networks of the Brain}}.\hskip 1em plus 0.5em minus
  0.4em\relax The MIT press, 10 2010. [Online]. Available:
  \url{https://doi.org/10.7551/mitpress/8476.003.0004}
\BIBentrySTDinterwordspacing

\bibitem{Sophie2007013}
S.~Achard and E.~Bullmore, ``Efficiency and cost of economical brain functional
  networks,'' \emph{PLoS computational biology}, vol.~3, p. e17, 03 2007.

\bibitem{Bullmore2009012}
E.~Bullmore and O.~Sporns, ``Complex brain networks: Graph theoretical analysis
  of structural and functional systems,'' \emph{Nature reviews Neuroscience},
  vol.~10, pp. 186--198, 2009.

\bibitem{liu2017detecting}
Q.~Liu, S.~Farahibozorg, C.~Porcaro, N.~Wenderoth, and D.~Mantini, ``Detecting
  large-scale networks in the human brain using high-density
  electroencephalography,'' \emph{Human brain mapping}, vol.~38, no.~9, pp.
  4631--4643, 2017.

\bibitem{Bassett200907}
D.~S. Bassett, E.~T. Bullmore, A.~Meyer-Lindenberg, J.~A. Apud, D.~R.
  Weinberger, and R.~Coppola, ``Cognitive fitness of cost-efficient brain
  functional networks,'' \emph{Proceedings of the National Academy of
  Sciences}, vol. 106, no.~28, pp. 11\,747--11\,752, 2009.

\bibitem{ZHU2012611}
\BIBentryALTinterwordspacing
X.~Zhu, X.~Wang, J.~Xiao, J.~Liao, M.~Zhong, W.~Wang, and S.~Yao, ``Evidence of
  a dissociation pattern in resting-state default mode network connectivity in
  first-episode, treatment-naive major depression patients,'' \emph{Biological
  Psychiatry}, vol.~71, no.~7, pp. 611--617, 2012, neural Circuitry of Mood.
  [Online]. Available:
  \url{https://www.sciencedirect.com/science/article/pii/S0006322311011036}
\BIBentrySTDinterwordspacing

\bibitem{LynnFCNCMN}
\BIBentryALTinterwordspacing
C.~W. Lynn and D.~S. Bassett, ``{The physics of brain network structure,
  function and control},'' \emph{Nature Reviews Physics}, vol.~1, pp. 318--332,
  05 2019. [Online]. Available: \url{https://doi.org/10.1038/s42254-019-0040-8}
\BIBentrySTDinterwordspacing

\bibitem{VANDENHEUVEL2010519}
\BIBentryALTinterwordspacing
M.~P. {van den Heuvel} and H.~E. {Hulshoff Pol}, ``Exploring the brain network:
  A review on resting-state fmri functional connectivity,'' \emph{European
  Neuropsychopharmacology}, vol.~20, no.~8, pp. 519--534, 2010. [Online].
  Available:
  \url{https://www.sciencedirect.com/science/article/pii/S0924977X10000684}
\BIBentrySTDinterwordspacing

\bibitem{ZHANG2012109}
\BIBentryALTinterwordspacing
Y.~Zhang, L.~Lin, C.-P. Lin, Y.~Zhou, K.-H. Chou, C.-Y. Lo, T.-P. Su, and
  T.~Jiang, ``Abnormal topological organization of structural brain networks in
  schizophrenia,'' \emph{Schizophrenia Research}, vol. 141, no.~2, pp.
  109--118, 2012. [Online]. Available:
  \url{https://www.sciencedirect.com/science/article/pii/S092099641200504X}
\BIBentrySTDinterwordspacing

\bibitem{LifeiDTI}
\BIBentryALTinterwordspacing
F.~Li, S.~Lui, L.~Yao, G.-J. Ji, W.~Liao, J.~A. Sweeney, and Q.~Gong,
  ``{Altered White Matter Connectivity Within and Between Networks in
  Antipsychotic-Naive First-Episode Schizophrenia},'' \emph{Schizophrenia
  Bulletin}, vol.~44, no.~2, pp. 409--418, 05 2017. [Online]. Available:
  \url{https://doi.org/10.1093/schbul/sbx048}
\BIBentrySTDinterwordspacing

\bibitem{ZUGMAN201589}
\BIBentryALTinterwordspacing
A.~Zugman, I.~Assunção, G.~Vieira, A.~Gadelha, T.~P. White, P.~P.~M.
  Oliveira, C.~Noto, N.~Crossley, P.~Mcguire, Q.~Cordeiro, S.~I. Belangero,
  R.~A. Bressan, A.~P. Jackowski, and J.~R. Sato, ``Structural covariance in
  schizophrenia and first-episode psychosis: An approach based on graph
  analysis,'' \emph{Journal of Psychiatric Research}, vol.~71, pp. 89--96,
  2015. [Online]. Available:
  \url{https://www.sciencedirect.com/science/article/pii/S0022395615002794}
\BIBentrySTDinterwordspacing

\bibitem{LiCMN2020}
\BIBentryALTinterwordspacing
Z.~Li, J.~Li, N.~Wang, and J.~Wang, ``Towards understanding interindividual
  differences in cortical morphological brain networks,'' \emph{bioRxiv}, 2020.
  [Online]. Available:
  \url{https://www.biorxiv.org/content/early/2020/12/23/2020.12.21.423884}
\BIBentrySTDinterwordspacing

\bibitem{WANG2021179}
\BIBentryALTinterwordspacing
C.~Wang, T.~Oughourlian, T.~A. Tishler, F.~Anwar, C.~Raymond, A.~D. Pham,
  A.~Perschon, J.~P. Villablanca, J.~Ventura, K.~L. Subotnik, K.~H.
  Nuechterlein, and B.~M. Ellingson, ``Cortical morphometric correlational
  networks associated with cognitive deficits in first episode schizophrenia,''
  \emph{Schizophrenia Research}, vol. 231, pp. 179--188, 2021. [Online].
  Available:
  \url{https://www.sciencedirect.com/science/article/pii/S0920996421001420}
\BIBentrySTDinterwordspacing

\bibitem{JiangFES2021}
Y.~Jiang, Y.~Wang, H.~Huang, H.~He, Y.~Tang, W.~Sun, L.~Xu, Y.~Wei, T.~Zhang,
  H.~Hu, J.~Wang, J.~Wang, C.~Luo, and D.~Yao, ``Antipsychotics effects on
  network-level reconfiguration of cortical morphometry in first-episode
  schizophrenia,'' \emph{Schizophrenia Bulletin}, 01 2021.

\bibitem{DSM-IV}
\BIBentryALTinterwordspacing
M.~Flaum and N.~C. Andreasen, ``{Diagnostic Criteria for Schizophrenia and
  Related Disorders: Options for DSM-IV},'' \emph{Schizophrenia Bulletin},
  vol.~17, no.~1, pp. 133--142, 01 1991. [Online]. Available:
  \url{https://doi.org/10.1093/schbul/17.1.133}
\BIBentrySTDinterwordspacing

\bibitem{JIANG2013263}
\BIBentryALTinterwordspacing
T.~Jiang, ``Brainnetome: A new -ome to understand the brain and its
  disorders,'' \emph{NeuroImage}, vol.~80, pp. 263--272, 2013, mapping the
  Connectome. [Online]. Available:
  \url{https://www.sciencedirect.com/science/article/pii/S1053811913003182}
\BIBentrySTDinterwordspacing

\bibitem{DESIKAN2006968}
\BIBentryALTinterwordspacing
R.~S. Desikan, F.~Ségonne, B.~Fischl, B.~T. Quinn, B.~C. Dickerson,
  D.~Blacker, R.~L. Buckner, A.~M. Dale, R.~P. Maguire, B.~T. Hyman, M.~S.
  Albert, and R.~J. Killiany, ``An automated labeling system for subdividing
  the human cerebral cortex on mri scans into gyral based regions of
  interest,'' \emph{NeuroImage}, vol.~31, no.~3, pp. 968--980, 2006. [Online].
  Available:
  \url{https://www.sciencedirect.com/science/article/pii/S1053811906000437}
\BIBentrySTDinterwordspacing

\bibitem{POTVIN201743}
\BIBentryALTinterwordspacing
O.~Potvin, L.~Dieumegarde, and S.~Duchesne, ``Freesurfer cortical normative
  data for adults using desikan-killiany-tourville and ex vivo protocols,''
  \emph{NeuroImage}, vol. 156, pp. 43--64, 2017. [Online]. Available:
  \url{https://www.sciencedirect.com/science/article/pii/S1053811917303269}
\BIBentrySTDinterwordspacing

\bibitem{DK2JAO}
C.~Jao, C.~Lau, L.~Lien, Y.~Tsai, K.~Chu, C.~Hsiao, J.~Yeh, and Y.~Wu,
  ``\BIBforeignlanguage{English}{Using fractal dimension analysis with the
  desikan–killiany atlas to assess the effects of normal aging on subregional
  cortex alterations in adulthood},'' \emph{\BIBforeignlanguage{English}{Brain
  Sciences}}, vol.~11, no.~1, pp. 1--17, Jan. 2021, funding Information: This
  research was funded by grants from Shin Kong Wu Ho-Su Memorial Hospital,
  Taipei, Taiwan (109GB006-3). The authors thank all the subjects who
  participated in this study, Wallace Academic Editing Company for editing this
  manuscript, and Mr. Yang-Ta Tseng for his effort in acquiring MRI data.
  Funding Information: Funding: This research was funded by grants from Shin
  Kong Wu Ho-Su Memorial Hospital, Taipei, Taiwan (109GB006-3). Publisher
  Copyright: {\textcopyright} 2021 by the authors. Licensee MDPI, Basel,
  Switzerland.

\bibitem{MeunierModular}
\BIBentryALTinterwordspacing
D.~Meunier, R.~Lambiotte, and E.~Bullmore, ``Modular and hierarchically modular
  organization of brain networks,'' \emph{Frontiers in Neuroscience}, vol.~4,
  2010. [Online]. Available:
  \url{https://www.frontiersin.org/article/10.3389/fnins.2010.00200}
\BIBentrySTDinterwordspacing

\bibitem{AssortativityNewman}
\BIBentryALTinterwordspacing
M.~E.~J. Newman, ``Mixing patterns in networks,'' \emph{Phys. Rev. E}, vol.~67,
  p. 026126, Feb 2003. [Online]. Available:
  \url{https://link.aps.org/doi/10.1103/PhysRevE.67.026126}
\BIBentrySTDinterwordspacing

\bibitem{GAT}
K.~S. Hosseini~SM, Hoeft~F, ``{GAT: a graph-theoretical analysis toolbox for
  analyzing between-group differences in large-scale structural and functional
  brain networks.}'' \emph{PLoS one}, vol.~7, no.~7, p. e40709, 2012.

\bibitem{2006UpperDensity}
M.~Kaiser, ``Nonoptimal component placement, but short processing paths, due to
  long-distance projections in neural systems,'' \emph{PLOS Computational
  Biology}, vol.~2, 2006.

\bibitem{BrainNetwork}
\BIBentryALTinterwordspacing
A.~Fornito, A.~Zalesky, and E.~T. Bullmore, Eds., \emph{Chapter 1 - An
  Introduction to Brain Networks}.\hskip 1em plus 0.5em minus 0.4em\relax San
  Diego: Academic Press, 2016, pp. 1--35. [Online]. Available:
  \url{https://www.sciencedirect.com/science/article/pii/B9780124079083000017}
\BIBentrySTDinterwordspacing

\bibitem{Schinack2015}
H.~G. Schnack, N.~E.~M. van Haren, R.~M. Brouwer, A.~Evans, S.~Durston, D.~I.
  Boomsma, R.~S. Kahn, and H.~E.~H. Pol, ``Changes in thickness and surface
  area of the human cortex and their relationship with intelligence,''
  \emph{Cereb Cortex}, vol.~25, pp. 1608--17, 2015.

\bibitem{2Tang2021}
Y.~Song, L.~Zou, J.~Zhao, X.~Zhou, Y.~Huang, H.~Qiu, H.~Han, Z.~Yang, X.~Li,
  X.~Tang, and J.~Chu, ``Whole brain volume and cortical thickness
  abnormalities in lson’s disease: a clinical correlation study,''
  \emph{Brain Imaging and Behavior}, vol.~15, pp. 1--10, 08 2021.

\bibitem{BAJAJ201836}
\BIBentryALTinterwordspacing
S.~Bajaj, A.~Raikes, R.~Smith, N.~S. Dailey, A.~Alkozei, J.~R. Vanuk, and W.~D.
  Killgore, ``The relationship between general intelligence and cortical
  structure in healthy individuals,'' \emph{Neuroscience}, vol. 388, pp.
  36--44, 2018. [Online]. Available:
  \url{https://www.sciencedirect.com/science/article/pii/S0306452218304834}
\BIBentrySTDinterwordspacing

\bibitem{CMNonADHD_autism}
\BIBentryALTinterwordspacing
R.~A.~I. Bethlehem, R.~Romero-Garcia, E.~Mak, E.~T. Bullmore, and
  S.~Baron-Cohen, ``{Structural Covariance Networks in Children with Autism or
  ADHD},'' \emph{Cerebral Cortex}, vol.~27, no.~8, pp. 4267--4276, 06 2017.
  [Online]. Available: \url{https://doi.org/10.1093/cercor/bhx135}
\BIBentrySTDinterwordspacing

\bibitem{SUI2015794}
\BIBentryALTinterwordspacing
J.~Sui, G.~D. Pearlson, Y.~Du, Q.~Yu, T.~R. Jones, J.~Chen, T.~Jiang,
  J.~Bustillo, and V.~D. Calhoun, ``In search of multimodal neuroimaging
  biomarkers of cognitive deficits in schizophrenia,'' \emph{Biological
  Psychiatry}, vol.~78, no.~11, pp. 794--804, 2015, schizophrenia:
  Glutamatergic Mechanisms of Cognitive Dysfunction and Treatment. [Online].
  Available:
  \url{https://www.sciencedirect.com/science/article/pii/S0006322315001274}
\BIBentrySTDinterwordspacing

\bibitem{CABALLERO20164}
A.~Caballero, R.~Granberg, and K.~Y. Tseng, ``Mechanisms contributing to
  prefrontal cortex maturation during adolescence,'' \emph{Neuroscience \&
  Biobehavioral Reviews}, vol.~70, pp. 4--12, 2016.

\bibitem{BENOIT2022}
\BIBentryALTinterwordspacing
L.~J. Benoit, S.~Canetta, and C.~Kellendonk, ``Thalamo-cortical development: A
  neurodevelopmental framework for schizophrenia,'' \emph{Biological
  Psychiatry}, 2022. [Online]. Available:
  \url{https://www.sciencedirect.com/science/article/pii/S0006322322010745}
\BIBentrySTDinterwordspacing

\bibitem{HANFORD2019129}
\BIBentryALTinterwordspacing
L.~C. Hanford, F.~Pinnock, G.~B. Hall, and R.~W. Heinrichs, ``Cortical
  thickness correlates of cognitive performance in cognitively-matched
  individuals with and without schizophrenia,'' \emph{Brain and Cognition},
  vol. 132, pp. 129--137, 2019. [Online]. Available:
  \url{https://www.sciencedirect.com/science/article/pii/S027826261830455X}
\BIBentrySTDinterwordspacing

\bibitem{JHUNG201335}
\BIBentryALTinterwordspacing
K.~Jhung, S.-H. Cho, J.-H. Jang, J.~Y. Park, D.~Shin, K.~R. Kim, E.~Lee, K.-H.
  Cho, and S.~K. An, ``Small-world networks in individuals at ultra-high risk
  for psychosis and first-episode schizophrenia during a working memory task,''
  \emph{Neuroscience Letters}, vol. 535, pp. 35--39, 2013. [Online]. Available:
  \url{https://www.sciencedirect.com/science/article/pii/S0304394012015467}
\BIBentrySTDinterwordspacing

\bibitem{WattsStrogatz}
D.~J. Watts and S.~H. Strogatz, ``Collective dynamics of "small-world"
  networks.'' \emph{Nature}, vol. 393, no.~1, pp. 440--442, 1998.

\bibitem{Small-worldNon}
\BIBentryALTinterwordspacing
M.~Rubinov, S.~A. Knock, C.~J. Stam, S.~Micheloyannis, A.~W. Harris, L.~M.
  Williams, and M.~Breakspear, ``Small-world properties of nonlinear brain
  activity in schizophrenia,'' \emph{Human Brain Mapping}, vol.~30, no.~2, pp.
  403--416, 2009. [Online]. Available:
  \url{https://onlinelibrary.wiley.com/doi/abs/10.1002/hbm.20517}
\BIBentrySTDinterwordspacing

\bibitem{Lynall2010022}
M.~E. Lynall, D.~Bassett, R.~Kerwin, P.~McKenna, M.~Kitzbichler,
  U.~Müller-Sedgwick, and E.~Bullmore, ``Functional connectivity and brain
  networks in schizophrenia,'' \emph{The Journal of neuroscience : the official
  journal of the Society for Neuroscience}, vol.~30, pp. 9477--87, 07 2010.

\bibitem{liu2008disrupted}
Y.~Liu, M.~Liang, Y.~Zhou, Y.~He, Y.~Hao, M.~Song, C.~Yu, H.~Liu, Z.~Liu, and
  T.~Jiang, ``Disrupted small-world networks in schizophrenia,'' \emph{Brain},
  vol. 131, no.~4, pp. 945--961, 2008.

\bibitem{ZHOU2021027}
\BIBentryALTinterwordspacing
H.~Zhou, L.~Shi, Y.~Shen, Y.~Fang, Y.~He, H.~Li, X.~Luo, E.~F. Cheung, and
  R.~C. Chan, ``Altered topographical organization of grey matter structural
  network in early-onset schizophrenia,'' \emph{Psychiatry Research:
  Neuroimaging}, vol. 316, p. 111344, 2021. [Online]. Available:
  \url{https://www.sciencedirect.com/science/article/pii/S0925492721000962}
\BIBentrySTDinterwordspacing

\bibitem{Wei2012017}
T.~Wei, X.~Liang, Y.~He, Y.-F. Zang, Z.~Han, A.~Caramazza, and Y.~Bi,
  ``Predicting conceptual processing capacity from spontaneous neuronal
  activity of the left middle temporal gyrus,'' \emph{The Journal of
  neuroscience : the official journal of the Society for Neuroscience},
  vol.~32, pp. 481--9, 01 2012.

\bibitem{Woodward018}
N.~Woodward, B.~Rogers, and S.~Heckers, ``Functional resting-state networks are
  differentially affected in schizophrenia.'' \emph{Schizophrenia research},
  vol. 130, pp. 86--93, 03 2011.

\bibitem{Zhao1998019}
J.~Zhao and D.~Yang, ``Research progress of cognitive function of
  schizophrenia,'' \emph{Chinese Journal of Psychiatry}, vol. 031, no. 001, pp.
  58--59, 1998.

\bibitem{KANG2018210}
\BIBentryALTinterwordspacing
S.~S. Kang, A.~W. MacDonald, M.~V. Chafee, C.-H. Im, E.~M. Bernat, N.~D.
  Davenport, and S.~R. Sponheim, ``Abnormal cortical neural synchrony during
  working memory in schizophrenia,'' \emph{Clinical Neurophysiology}, vol. 129,
  no.~1, pp. 210--221, 2018. [Online]. Available:
  \url{https://www.sciencedirect.com/science/article/pii/S1388245717311197}
\BIBentrySTDinterwordspacing

\bibitem{YAO2021187}
\BIBentryALTinterwordspacing
R.~Yao, J.~Xue, P.~Yang, Q.~Wang, P.~Gao, X.~Yang, H.~Deng, S.~Tan, and H.~Li,
  ``Dynamic changes of brain networks during working memory tasks in
  schizophrenia,'' \emph{Neuroscience}, vol. 453, pp. 187--205, 2021. [Online].
  Available:
  \url{https://www.sciencedirect.com/science/article/pii/S0306452220307235}
\BIBentrySTDinterwordspacing

\bibitem{Liu2004020}
D.~Liu, Y.~Xu, and L.~Zheng, ``Functional magnetic resonance imaging of
  backward digit span task in first-episode schizophrenic patients,''
  \emph{Shanghai Archives of Psychiatry}, vol. 016, no. 005, 2004.

\bibitem{HOLMES1989118}
\BIBentryALTinterwordspacing
G.~L. Holmes, ``The prefrontal cortex: Anatomy, physiology, and neuropsychology
  of the frontal lobe (2nd ed.): by joaquin m. fuster. new york: Raven press,
  1988, 269 pp.'' \emph{Journal of Epilepsy}, vol.~2, no.~2, p. 118, 1989.
  [Online]. Available:
  \url{https://www.sciencedirect.com/science/article/pii/0896697489900492}
\BIBentrySTDinterwordspacing

\bibitem{Gong55}
\BIBentryALTinterwordspacing
J.~Gong, J.~Wang, X.~Luo, G.~Chen, H.~Huang, R.~Huang, L.~Huang, and Y.~Wang,
  ``Abnormalities of intrinsic regional brain activity in first-episode and
  chronic schizophrenia: a meta-analysis of resting-state functional mri,''
  \emph{Journal of Psychiatry and Neuroscience}, vol.~45, no.~1, pp. 55--68,
  2020. [Online]. Available: \url{https://www.jpn.ca/content/45/1/55}
\BIBentrySTDinterwordspacing

\bibitem{HARRISON2008421}
\BIBentryALTinterwordspacing
P.~J. Harrison, ``Neuropathology of schizophrenia,'' \emph{Psychiatry}, vol.~7,
  no.~10, pp. 421--424, 2008, schizophrenia Part 1 of 2. [Online]. Available:
  \url{https://www.sciencedirect.com/science/article/pii/S1476179308001560}
\BIBentrySTDinterwordspacing

\bibitem{LIANG2020026}
\BIBentryALTinterwordspacing
S.~Liang, W.~Deng, X.~Li, Q.~Wang, A.~J. Greenshaw, W.~Guo, X.~Kong, M.~Li,
  L.~Zhao, Y.~Meng, C.~Zhang, H.~Yu, X.~min Li, X.~Ma, and T.~Li, ``Aberrant
  posterior cingulate connectivity classify first-episode schizophrenia from
  controls: A machine learning study,'' \emph{Schizophrenia Research}, vol.
  220, pp. 187--193, 2020. [Online]. Available:
  \url{https://www.sciencedirect.com/science/article/pii/S0920996420301286}
\BIBentrySTDinterwordspacing

\bibitem{HAZNEDAR2004025}
\BIBentryALTinterwordspacing
M.~Haznedar, M.~S. Buchsbaum, E.~A. Hazlett, L.~Shihabuddin, A.~New, and L.~J.
  Siever, ``Cingulate gyrus volume and metabolism in the schizophrenia
  spectrum,'' \emph{Schizophrenia Research}, vol.~71, no.~2, pp. 249--262,
  2004. [Online]. Available:
  \url{https://www.sciencedirect.com/science/article/pii/S092099640400088X}
\BIBentrySTDinterwordspacing

\bibitem{archpsyc1991028}
\BIBentryALTinterwordspacing
S.~E. Arnold, B.~T. Hyman, G.~W. Van~Hoesen, and A.~R. Damasio, ``{Some
  Cytoarchitectural Abnormalities of the Entorhinal Cortex in Schizophrenia},''
  \emph{Archives of General Psychiatry}, vol.~48, no.~7, pp. 625--632, 07 1991.
  [Online]. Available:
  \url{https://doi.org/10.1001/archpsyc.1991.01810310043008}
\BIBentrySTDinterwordspacing

\bibitem{bleuler1911dementia}
E.~Bleuler, \emph{Dementia praecox oder Gruppe der Schizophrenien}.\hskip 1em
  plus 0.5em minus 0.4em\relax Deuticke, 1911, vol.~4.

\bibitem{4YUAN2022109441}
\BIBentryALTinterwordspacing
J.~Yuan, X.~Ran, K.~Liu, C.~Yao, Y.~Yao, H.~Wu, and Q.~Liu, ``Machine learning
  applications on neuroimaging for diagnosis and prognosis of epilepsy: A
  review,'' \emph{Journal of Neuroscience Methods}, vol. 368, p. 109441, 2022.
  [Online]. Available:
  \url{https://www.sciencedirect.com/science/article/pii/S0165027021003769}
\BIBentrySTDinterwordspacing

\bibitem{CANNON202212}
\BIBentryALTinterwordspacing
T.~D. Cannon, ``Psychosis, schizophrenia, and states vs. traits,''
  \emph{Schizophrenia Research}, vol. 242, pp. 12--14, 2022, re-Inventing
  Schizophrenia: Updating the Construct. [Online]. Available:
  \url{https://www.sciencedirect.com/science/article/pii/S0920996421004850}
\BIBentrySTDinterwordspacing

\end{thebibliography}

\end{document}